\documentclass[twocolumn]{aastex61}
\pdfoutput=1 %for arXiv submission
\usepackage{amsmath,amstext}
\usepackage[T1]{fontenc}
\usepackage{apjfonts} 
\usepackage[figure,figure*]{hypcap}
\usepackage{textcomp}
\usepackage{color}
\usepackage{booktabs}
\usepackage{catoptions}
\usepackage{amsmath}
\usepackage[maxfloats=40]{morefloats}
\usepackage{placeins}
\usepackage[normalem]{ulem}

\newcommand{\Rom}[1]{\uppercase\expandafter{\romannumeral #1\relax}}

\def \pext {$P_{ext}$}
\def \pemit {$P_{emit}$}
\def \jyb {Jy~beam$^{-1}$}

\begin{document}
\submitjournal{\apj}
\accepted{March 25, 2019}

\AuthorCallLimit=500
\title{JCMT BISTRO survey: Magnetic Fields within the Hub-Filament Structure in IC 5146}

\author[0000-0002-6668-974X]{Jia-Wei Wang}
\affiliation{Institute of Astronomy and Department of Physics, National Tsing Hua University, Hsinchu 30013, Taiwan}

\author[0000-0001-5522-486X]{Shih-Ping Lai}
\affiliation{Institute of Astronomy and Department of Physics, National Tsing Hua University, Hsinchu 30013, Taiwan}

\author[0000-0003-4761-6139]{Chakali Eswaraiah} 
\affiliation{CAS Key Laboratory of FAST, National Astronomical Observatories, Chinese Academy of Sciences, People's Republic of China}
\affiliation{Institute of Astronomy and Department of Physics, National Tsing Hua University, Hsinchu 30013, Taiwan}

\author[0000-0002-8557-3582]{Kate Pattle}
\affiliation{Institute of Astronomy and Department of Physics, National Tsing Hua University, Hsinchu 30013, Taiwan}

\author[0000-0002-9289-2450]{James Di Francesco} 
\affiliation{NRC Herzberg Astronomy and Astrophysics, 5071 West Saanich Road, Victoria, BC V9E 2E7, Canada}
\affiliation{Department of Physics and Astronomy, University of Victoria, Victoria, BC V8P 5C2, Canada}

\author[0000-0002-6773-459X]{Doug Johnstone}
\affiliation{NRC Herzberg Astronomy and Astrophysics, 5071 West Saanich Road, Victoria, BC V9E 2E7, Canada}
\affiliation{Department of Physics and Astronomy, University of Victoria, Victoria, BC V8P 5C2, Canada}

\author[0000-0003-2777-5861]{Patrick M. Koch}
\affiliation{Academia Sinica Institute of Astronomy and Astrophysics, P.O. Box 23-141, Taipei 10617, Taiwan}

\author[0000-0002-5286-2564]{Tie Liu}
\affiliation{Korea Astronomy and Space Science Institute, 776 Daedeokdae-ro, Yuseong-gu, Daejeon 34055, Republic of Korea}
\affiliation{East Asian Observatory, 660 N. A`oh\={o}k\={u} Place, University Park, Hilo, HI 96720, USA}

\author[0000-0002-6510-0681]{Motohide Tamura}
\affiliation{National Astronomical Observatory of Japan, National Institutes of Natural Sciences, Osawa, Mitaka, Tokyo 181-8588, Japan}
\affiliation{Department of Astronomy, Graduate School of Science, The University of Tokyo, 7-3-1 Hongo, Bunkyo-ku, Tokyo 113-0033, Japan}
\affiliation{Astrobiology Center, National Institutes of Natural Sciences, 2-21-1 Osawa, Mitaka, Tokyo 181-8588, Japan}

\author[0000-0003-0646-8782]{Ray S. Furuya} 
\affiliation{Tokushima University, Minami Jousanajima-machi 1-1, Tokushima 770-8502, Japan}
\affiliation{Institute of Liberal Arts and Sciences Tokushima University, Minami Jousanajima-machi 1-1, Tokushima 770-8502, Japan}

% ...

\begin{abstract}
We present the 850 $\mu$m polarization observations toward the IC5146 filamentary cloud taken using the Submillimetre Common-User Bolometer Array 2 (SCUBA-2) and its associated polarimeter (POL-2), mounted on the James Clerk Maxwell Telescope (JCMT), as part of the B-fields In STar forming Regions Observations (BISTRO). This work is aimed at revealing the magnetic field morphology within a core-scale ($\lesssim 1.0$ pc) hub-filament structure (HFS) located at the end of a parsec-scale filament. To investigate whether or not the observed polarization traces the magnetic field in the HFS, we analyze the dependence between the observed polarization fraction and total intensity using a Bayesian approach with the polarization fraction described by the Rice likelihood function, which can correctly describe the probability density function (PDF) of the observed polarization fraction for low signal-to-noise ratio (SNR) data. We find a power-law dependence between the polarization fraction and total intensity with an index of 0.56 in $A_V\sim$ 20--300 mag regions, suggesting that the dust grains in these dense regions can still be aligned with magnetic fields in the IC5146 regions. Our polarization maps reveal a curved magnetic field, possibly dragged by the contraction along the parsec-scale filament. We further obtain a magnetic field strength of 0.5$\pm$0.2 mG toward the central hub using the Davis-Chandrasekhar-Fermi method, corresponding to a mass-to-flux criticality of $\sim$ $1.3\pm0.4$ and an Alfv\'{e}nic Mach number of $<$0.6. These results suggest that gravity and magnetic field is currently of comparable importance in the HFS, and turbulence is less important.

\end{abstract}

\section{Introduction}\label{sec:intro}
Observations over the last few decades have revealed that stars predominately form within magnetized and turbulent molecular clouds \citep{cr12}. How magnetic fields regulate star formation, however, is still poorly understood. Theoretical works have suggested that magnetic fields could be important in supporting molecular clouds, suppressing the star formation rate \citep[e.g.,][]{na08,pr08}, and removing angular momentum \citep[e.g.,][]{me85,mo86}. Nonetheless, measurements of magnetic field morphologies and strength are still too rare to test these theories \citep[and references therein]{ta87,ta15,kw15}.

Recently, much attention has been drawn to filamentary molecular clouds, which are suggested as the key progenitors of star formation \citep{an10}. \citet{li13} found that the intercloud media magnetic fields, traced by optical polarimetry, were often oriented either parallel or perpendicular to the filamentary clouds. Based on the recent \textit{Planck} data, \citet{pl16} showed that the relative orientations between magnetic fields and filamentary density structure changed systematically from parallel to filaments in low column density areas to perpendicular in high column density areas, with a switch point at $N_{H}\sim 10^{21.7} cm^{-2}$. The observed alignment between magnetic fields and filaments is consistent with theoretical works suggesting that magnetic fields play an important role in guiding the gravitational or turbulence-driven contraction and also supporting filaments from contraction along the filament major axis \citep[e.g.,][]{na08,bu13,in15}. 

Within dense filamentary clouds, morphological configurations named ``Hub-Filament structures'' (HFSs) are commonly seen. Such structures consist of a central dense hub ($N_{H_2}>10^{22}~cm^{-2}$) with several converging filaments surrounding the hub \citep{my09,li14}. The central hubs of HFSs often host most of the star formation in a filamentary cloud, and hence are the potential sites for cluster formation. Observations found that converging filaments connecting HFSs often have similar orientations and spacings \citep{my09}. Polarization observations toward the HFS G14.225-0.506 found that its converging filaments are perpendicular to the large-scale magnetic fields \citep{bu13,sa16}. To explain these features, theoretical works have suggested that HFSs are formed via layer fragmentation of clouds threaded by magnetic fields. In this model, the local densest regions collapse quickly and form dense hubs, then the surrounding material tends to fragment along magnetic field lines and become parallel layers, since the gravitational instability grows faster along the magnetic fields \citep{na98,my09,va14}.

Kinematic analyses have shown that the surrounding filaments within HFSs might indeed be infalling material \citep[e.g.,][]{liu15,ju17,yu17}, attributed either to accretion flows attracted by the dense hubs \citep[e.g.,][]{fr13,ki13} or to gravitationally collapsing filaments \citep[e.g.,][]{po11,po12}. As an example, spiral arm-like converging filaments with significant velocity gradients were revealed in G33.92+0.11 using ALMA, supporting the idea that these filaments were eccentric accretion flows preserving high angular momentum and were previously fragmented from a rotating clump \citep{liu12,liu15}. In this point of view, the formation of HFSs is dominated by gravity, and magnetic fields are merely dragged by the accretion flows. Therefore, the magnetic field morphologies are parallel to the converging filaments and different from the larger scale magnetic fields, which have been seen in NGC 6334 V via SMA polarimetry \citep{ju17}. Since most of the current HFS formation models were based on the large-scale magnetic field morphologies, it is still challenging to explain how the observed infalling features evolve at smaller scale. Hence, more observations on the scale of HFS are essential to complete evolutionary models.

Submillimeter dust polarization is commonly used to measure the magnetic field morphology. Whether or not submillimeter polarization can really trace the dust grains in dense clouds, however, is still in debate. Current radiative torque dust alignment (RATs) theory \citep{la97,la07,ho09} suggests that dust grains in high-extinction regions cannot be efficiently aligned with magnetic fields due to the lack of radiation fields. Observationally, polarization efficiency ($PE$), defined as a ratio of absorption polarization percentage to visual extinction $A_V$, is commonly used to evaluate whether or not dust grains within clouds could be aligned. Past observations found that the polarization efficiency in high-extinction regions decreases with $A_V$ by a power-law index of -1, indicating that the polarization only comes from the surface of the clouds \citep[e.g.,][]{jo15,an15}, and also support the prediction of the RATs theory. Nevertheless, some observations do show flatter $PE$--$A_V$ relations \citep[e.g.,][]{jo16,wa17}, suggesting that the dust grains in dense regions do contribute polarization. It is still unclear what mechanism can efficiently align dust grains in high-extinction regions and what environment the mechanism requires. More measurements of polarization efficiency in different environments are needed to settle this debate.

The IC5146 system is a nearby star-forming region in Cygnus, consisting of an H\Rom{2} region, known as the Cocoon Nebula, and a long dark cloud extending from the H\Rom{2} region. The distance of the IC5146 cloud is ambiguous. \citet{ha08} estimated a distance of 950 pc based on the comparison of zero-age main sequence among the Orion Nebula Cluster and the B-type members of IC5146. In contrast, \citet{la99} derived a distance of $460\substack{+40 \\ -60}$\, pc by comparing the number of foreground stars, identified by near-infrared extinction measurements, to those expected from galactic models. \citet{dz18} estimated a distance of $813\pm106$ pc based on the Gaia second data release \citep{ga18} parallax measurements toward the embedded young stellar objects (YSOs) within the Cocoon Nebula. In this paper, we assume a default distance of $813\pm106$ pc for consistency.

The \textit{Herschel} Gould Belt Survey \citep{an10} revealed a complex network of filaments within the IC5146 dark clouds \citep{ar11}: several diffuse sub-filaments extend from its main filamentary structures, and two HFSs are located at the ends of the main filaments. The main filament is a known active star-forming region, where more than 200 YSOs have been identified with $Spitzer$ \citep{ha08,du15}. The variety of filamentary features in the IC5146 system suggests it as an ideal target for investigating the formation and evolution of these filaments \citep{jo17}. \citet{wa17} (hereafter WLE17) measured the optical and near-infrared starlight polarization across the whole IC5146 cloud, and showed that the large-scale magnetic fields are uniform and perpendicular to the main filaments, suggesting that the large-scale filaments were formed under strong magnetic field condition. Since the large-scale magnetic fields have been well probed, the IC5146 system is an excellent target to perform further submillimeter polarimetry to reveal the role of the magnetic field to smaller scales.

In this paper, we report the 850 $\mu$m polarization observations toward the brightest HFS in the IC5146 system taken with Submillimetre Common-User Bolometer Array 2 (SCUBA-2, \citealt{ho13}) and its associated polarimeter (POL-2, \citealt{fr16,ba19}), mounted on the James Clerk Maxwell Telescope (JCMT), as part of the B-fields In STar forming Regions Observations (BISTRO) \citep{war17,kw18,so18,pa18}.
The target HFS has a physical size less than $\sim$ 1.0 pc and a total mass of $\sim$ 100 $M_{\sun}$ \citep{ha08}, lower than the commonly seen parsec-scale HFS such as NGC1333 or IC348, and thus hereafter we named our target as ``core-scale HFS'' to distinguish it from other HFSs with much larger physical scale. In \autoref{sec:obs}, we address the details of our observations and data reduction. In \autoref{sec:results}, we discuss the magnetic field morphology revealed by the polarization map, and we present an analysis of the dependency between polarization efficiency and $A_V$, the magnetic field morphology, and magnetic strength. Our interpretations of the observed polarization data are discussed in \autoref{sec:discussion}. A summary of our conclusions is given in \autoref{sec:summary}.

\section{Observations}\label{sec:obs}
\subsection{Data Acquisition and Reduction Techniques}\label{sec:data}
\begin{figure*}
\includegraphics[width=\textwidth]{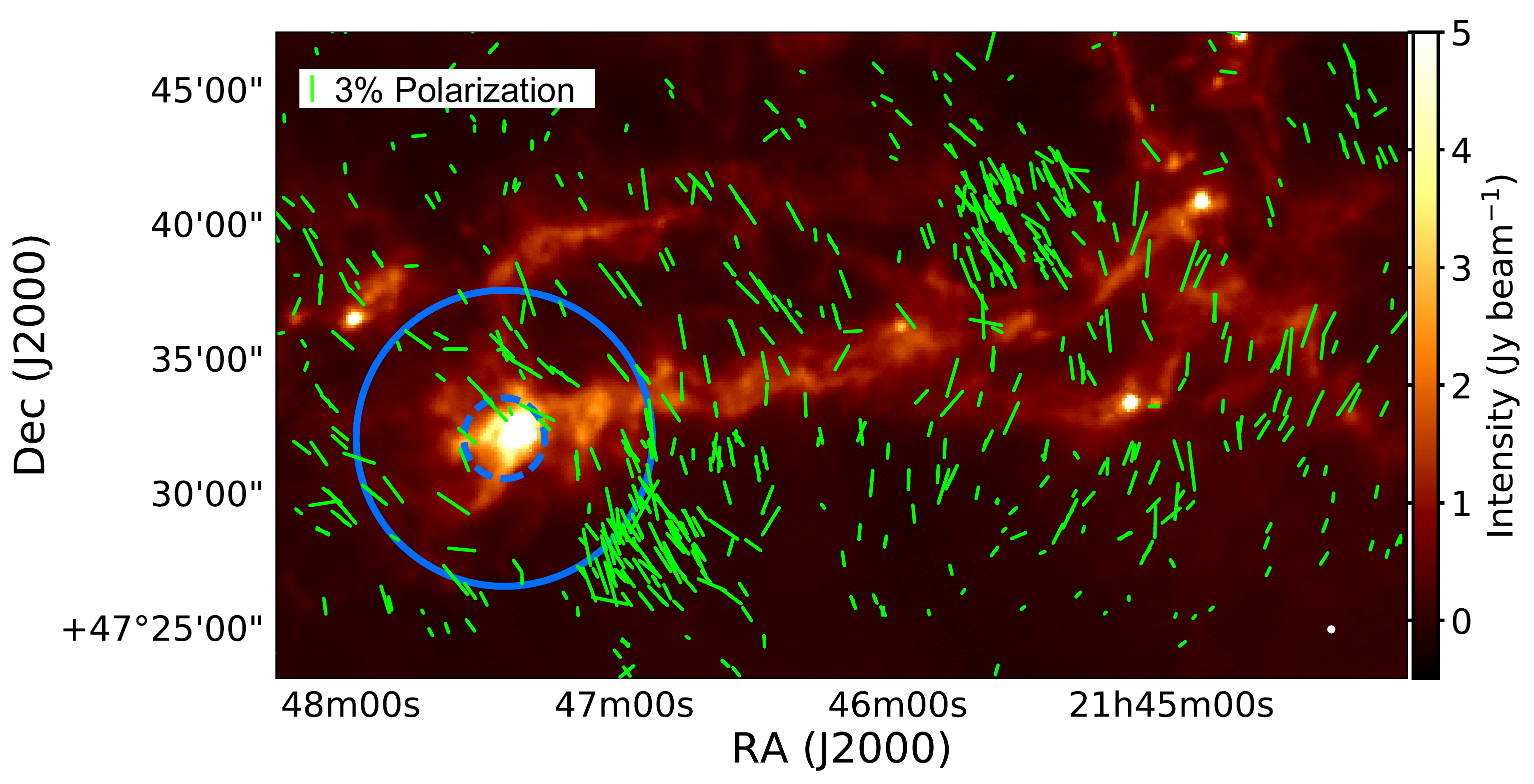}
\caption{The IC5146 field observed in BISTRO overlaid on the \textit{Herschel} 250 $\mu$m image. The blue solid circle represents the field of view of the POL-2 850 $\mu$m polarimetry observations, and the dashed circle indicates the inner 3\arcmin\ region with the best sensitivity. The green vectors show the optical and infrared polarization measurements \citep{wa17}. The white circle at the bottom right corner shows the $Herschel$ 250 $\mu$m FWHM beam size. We note that the Cocoon Nebula is about 1\degr\ to the east of this field.}\label{fig:field}
\end{figure*}

Our polarimetric continuum observations toward the IC5146 dark cloud system were carried out between 2016 May and 2017 April. The observed field targeted the brightest HFS located at the eastern end of the IC5146 main filament, as shown in \autoref{fig:field}. We performed 20 sets of 40-minute observations toward the IC5146 region with $\tau_{225 GHz}$ ranging from 0.04 to 0.07.

The POL-2 observations were made using POL-2 DAISY scan mode \citep{fr16,ba19}, producing a fully sampled circular region of 11 arcmin diameter. Within the DAISY map, the noise is uniform and lowest in the central 3 arcmin diameter region, and increases to the edge of the map. The POL-2 data were simultaneously taken at 450 $\mu$m with a resolution of 9.6 arcsec and at 850 $\mu$m with a resolution of 14.1 arcsec. The 450 $\mu$m data are not reported in this paper since the 450 $\mu$m instrumental polarization model has been only recently commissioned.

The IC5146 data were reduced in a three-stage process using $pol2map$\footnote{http://starlink.eao.hawaii.edu/docs/sc22.pdf}, a script recently added to the SCUBA-2 mapmaking routine \textsc{smurf} \citep{be05, ch13}.  

In the first stage, the raw bolometer timestreams for each observation are converted into separate Stokes $Q$, $U$, and $I$ timestreams using the process $calcqu$.

In the second stage, an initial Stokes $I$ map is created from the $I$ timestream from each observation using the iterative map-making routine $makemap$. For each reduction, areas of astrophysical emission are defined using a signal-to-noise-ratio (SNR) based mask determined iteratively by $makemap$. Areas outside this masked region are set to zero after each iteration until the final iteration of $makemap$ (see \citealt{ma15} for a detailed description of the role of masking in SCUBA-2 data reduction). Convergence is reached when successive iterations of the mapmaker produce pixel values which differ by $< 5$\% on average.  Each map is compared to the first map in the sequence to determine a set of relative pointing corrections. The individual $I$ maps are then coadded to produce an initial $I$ map of the region.  

In the third stage, the final Stokes $I,$ $Q$, and $U$ maps are created. The initial $I$ map described above is used to generate a fixed SNR-based mask for all further iterations of \textit{makemap}. The pointing corrections determined in the second stage are applied during the map-making process.  In this stage, $skyloop$, a variant mode of \textit{makemap}, is invoked.  In this mode, rather than each observation being reduced consecutively as is the standard method, one iteration of the mapmaker is performed on each of the observations. At the end of each iteration, all the maps created are coadded. The coadded maps created after successive iterations are compared, and when these coadded maps on average vary by $<5$\% between successive iterations, convergence is reached.  Using $skyloop$ typically improves the mapmaker's ability to recover faint extended structure, at the expense of additional memory usage and processing time.  The mapmaker was run three times successively to produce the output $I$, $Q$, and $U$ maps from their respective timestreams.  The $Q$ and $U$ data were corrected for instrumental polarization (IP) using the final output $I$ map and the latest IP model (January 2018) \citep{fr16, fr18}.

In all $pol2map$ instances of $makemap$, the polarized sky background is estimated by doing a principal component analysis (PCA) of the $I$, $Q$, and $U$ timestreams to identify components that are common to multiple bolometers. In the first run of \textit{makemap} (stage 2), the 50 most correlated components are removed at each iteration. In the second run (stage 3), 150 components are removed at each iteration, resulting in smaller changes in the map between iterations and lower noise in the final map.

The output $I$, $Q$, and $U$ maps were calibrated in units of Jy/beam, using a flux conversion factor (FCF) of 725 Jy/pW -- the standard 850$\mu$m SCUBA-2 FCF multiplied by 1.35 to account for additional losses due to POL-2 (\citealt{de13}, \citealt{fr16}).

Finally, a polarization vector catalogue was created from the coadded Stokes $I$, $Q$, and $U$ maps. To improve the sensitivity, we binned the coadded Stokes $I$, $Q$, and $U$ maps into 12\arcsec\ pixels, and the binned data reached rms noise levels of 1.1 m\jyb\ for Stokes $Q$ and $U$. 

We calculated the polarization fractions and orientations in the 12\arcsec\ pixel map. We debiased the former with the asymptotic estimator \citep{wa74} as
\begin{equation}\label{eq:debias}
P=\frac{1}{I}\sqrt{(U^2+Q^2)-\frac{1}{2}(\delta Q^2+\delta U^2)},
\end{equation}
where $P$ is the debiased polarization percentage, and $I$, $Q$, $U$, $\delta I$, $\delta Q$, and $\delta U$ are the Stokes $I$, $Q$, $U$, and their uncertainties. The uncertainty of polarization fraction was estimated using
\begin{equation}\label{eq:eP}
\delta P=\sqrt{\frac{(Q^2\delta Q^2+U^2\delta U^2)}{I^2(Q^2+U^2)} + \frac{\delta I^2(Q^2+U^2)}{I^4}}.
\end{equation}
The polarization position angle ($PA$) was calculated as:
\begin{equation}\label{eq:PA}
PA=\frac{1}{2}\tan^{-1}(\frac{U}{Q}),
\end{equation}
and its corresponding uncertainties were estimated using:
\begin{equation}\label{eq:ePA}
\delta PA=\frac{1}{2}\sqrt{\frac{(Q^2\delta U^2+U^2\delta Q^2)}{(Q^2+U^2)^2}} .
\end{equation}
The magnetic field orientations used in this paper were assumed to be $PA +90\degr$.

\subsection{CO Contamination}\label{sec:CO}
The SCUBA-2 850 $\mu$m waveband covers the wavelength of the CO (J=3-2) rotational line, and thus our measured continuum flux could be affected by CO line emission \citep[e.g.,][]{dr12,co16}. Furthermore, the CO (J=3-2) rotational line is known to be polarized via the Goldreich--Kylafis effect \citep{go81,go82}. For example, the typical polarization fraction of CO (J=3-2) could be $\lesssim$ 3\% in dense clouds and outflows \citep{ch16}, calculated using the formulation in \citet{de84} and \citet{co05}. The polarization angle of CO line is either parallel or perpendicular to the magnetic fields depending on optical depth and the relative angle between magnetic fields and gas velocity fields \citep{co05}. If a typical polarization fraction of 2\% for CO (J=3-2) is assumed, the polarized intensity from the line would be only 0.02--0.14\% of the total 850 $\mu$m flux, which is insignificant compared to the uncertainties of polarization, $\gtrsim 0.2$--$0.5\%$, in the central hub.

The CO contamination in total intensity might also decrease the observed polarization fraction. \citet{jo17} calculated the fraction of CO (J=3-2) line emission to the total flux in the JCMT 850 $\mu$m waveband toward several clumps in the IC5146 system. The fraction of CO (J=3-2) to total flux in our target region is mostly $\sim$ 1--3\%, but a higher fraction of $\sim$ 7\% was found in the central hub. Hence, the CO contamination would reduce the measured polarization fraction by a factor of 1--7\%. Nevertheless, this effect is insignificant to our analysis since the SNRs of our polarization detections are typically only $\sim$2-4.

\section{Results and Analysis}\label{sec:results}
\begin{figure*}
\includegraphics[width=\textwidth]{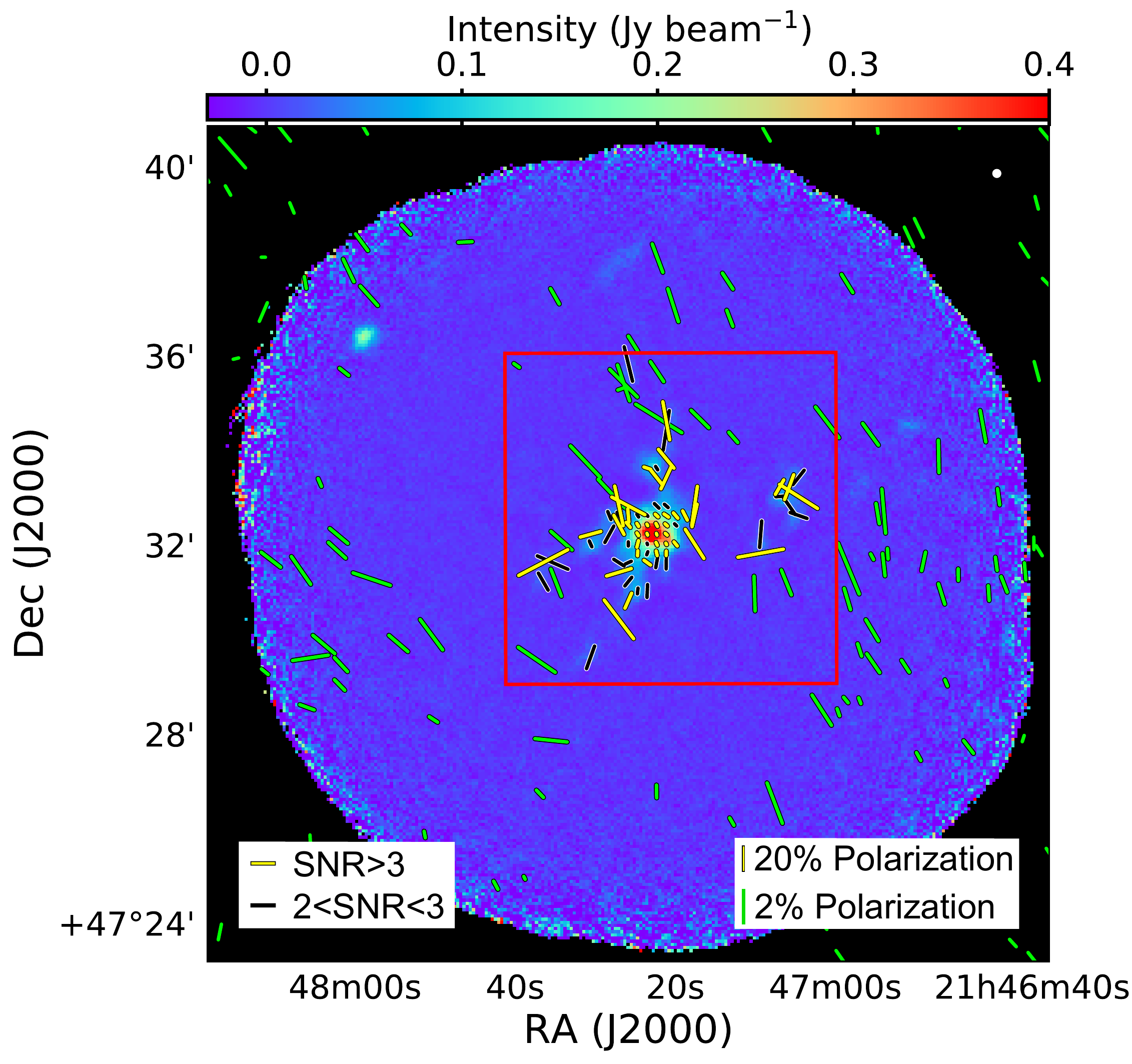}
\caption{B-field orientation map sampled on a 12\arcsec\ grid shown on the 850 $\mu$m dust continuum map, sampled on a 4\arcsec\ grid, of IC5146 region. The vectors are selected by $I/\delta I>10$ and $P/\delta P>2$, and rotated by 90\degr\ to represent magnetic field orientations. The yellow and black vectors show the greater-than-3$\sigma$ and 2--3$\sigma$ polarization detections. The green vectors represent the H-band starlight polarization. The white circle at the top right corner shows the POL-2 850 $\mu$m beam size of 14 arcsec. The zoom-in to the red box is shown in \autoref{fig:pmap}}\label{fig:pmap_all}
\end{figure*}

\begin{figure*}
\includegraphics[width=\textwidth]{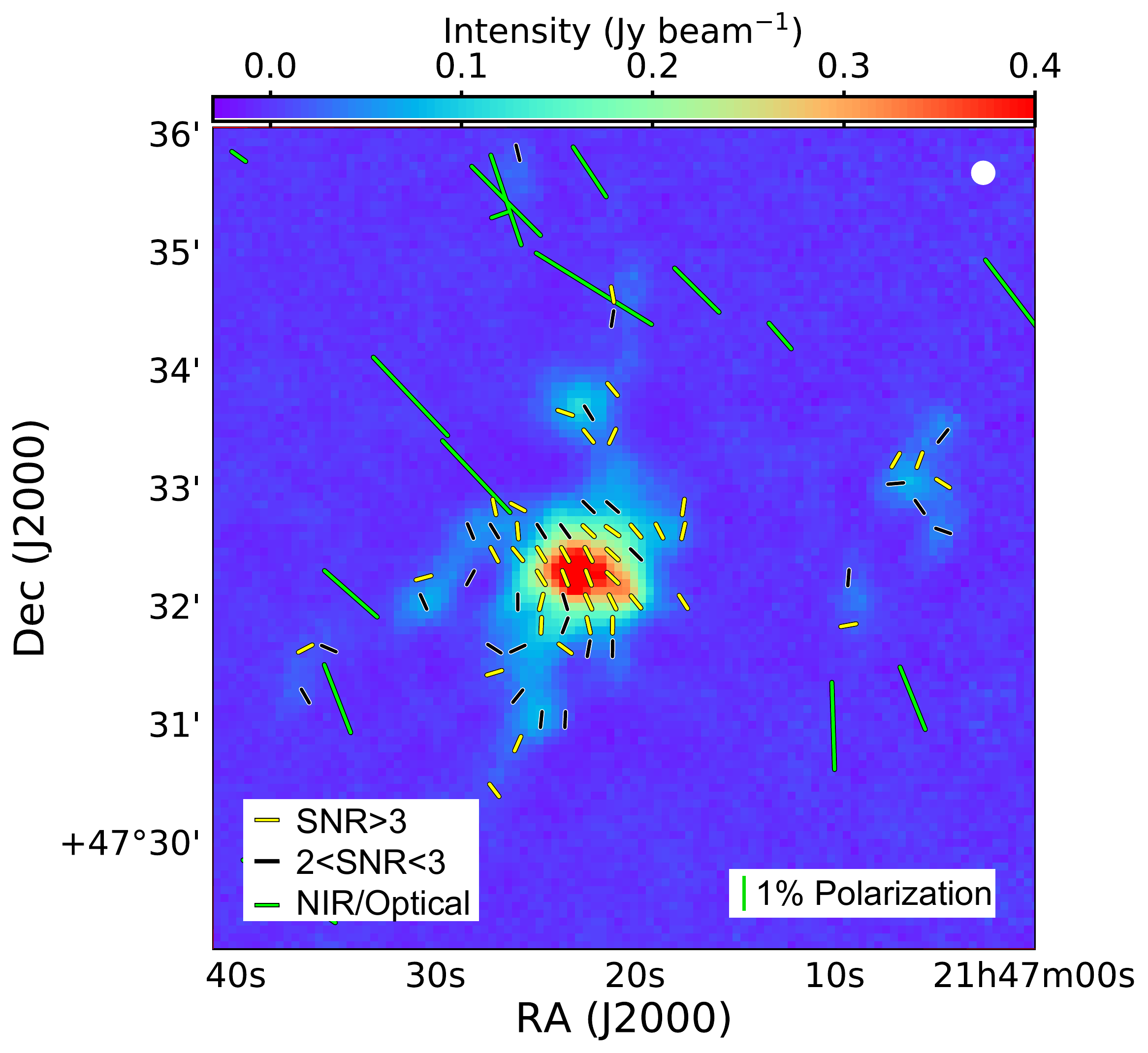}
\caption{Same as \autoref{fig:pmap_all}, but zoom into the HFS region. The vectors are plotted with constant lengths to emphasize their orientations.}\label{fig:pmap}
\end{figure*}

\subsection{Magnetic Field Morphology}\label{sec:pmap}
We show the observed magnetic field orientations traced by POL-2 850 $\mu$m polarization, with pixel size of 12\arcsec, overlaid on the Stokes $I$ map, with pixel size of 4\arcsec, in \autoref{fig:pmap_all}. We selected the 139 vectors with $I/\delta I>10$ to ensure that the selected data are associated with the target core-scale HFS. \citet{mo15a} suggest that the uncertainty in Stokes $I$ may enhance the bias in polarization fraction for data with $I/\delta I<10$, and thus our $I/\delta I>10$ selection criterion could exclude these biased data. Among these $I/\delta I>10$ data, 30 of them have $2<P/\delta P<3$ and 42 of them have $P/\delta P> 3$. In order to better probe the magnetic fields, we further added the $P/\delta P>2$ criterion to exclude the samples with higher uncertainties in $PA$, and the final selected samples have a maximum $\delta PA$ of 12.7\degr\ and a mean $\delta PA$ of 8.5\degr. \autoref{fig:pmap} shows the zoom-in polarization map toward the HFS and our final selected samples. We note that the CO contamination in Stokes $I$ only has insignificant effect on our sample selection. If assuming the CO contamination in total intensity is 7\% everywhere, as the worst case, the number of $P/\delta P>2$ vectors would only decrease to 68 from 72.

\begin{figure}
\includegraphics[width=\columnwidth]{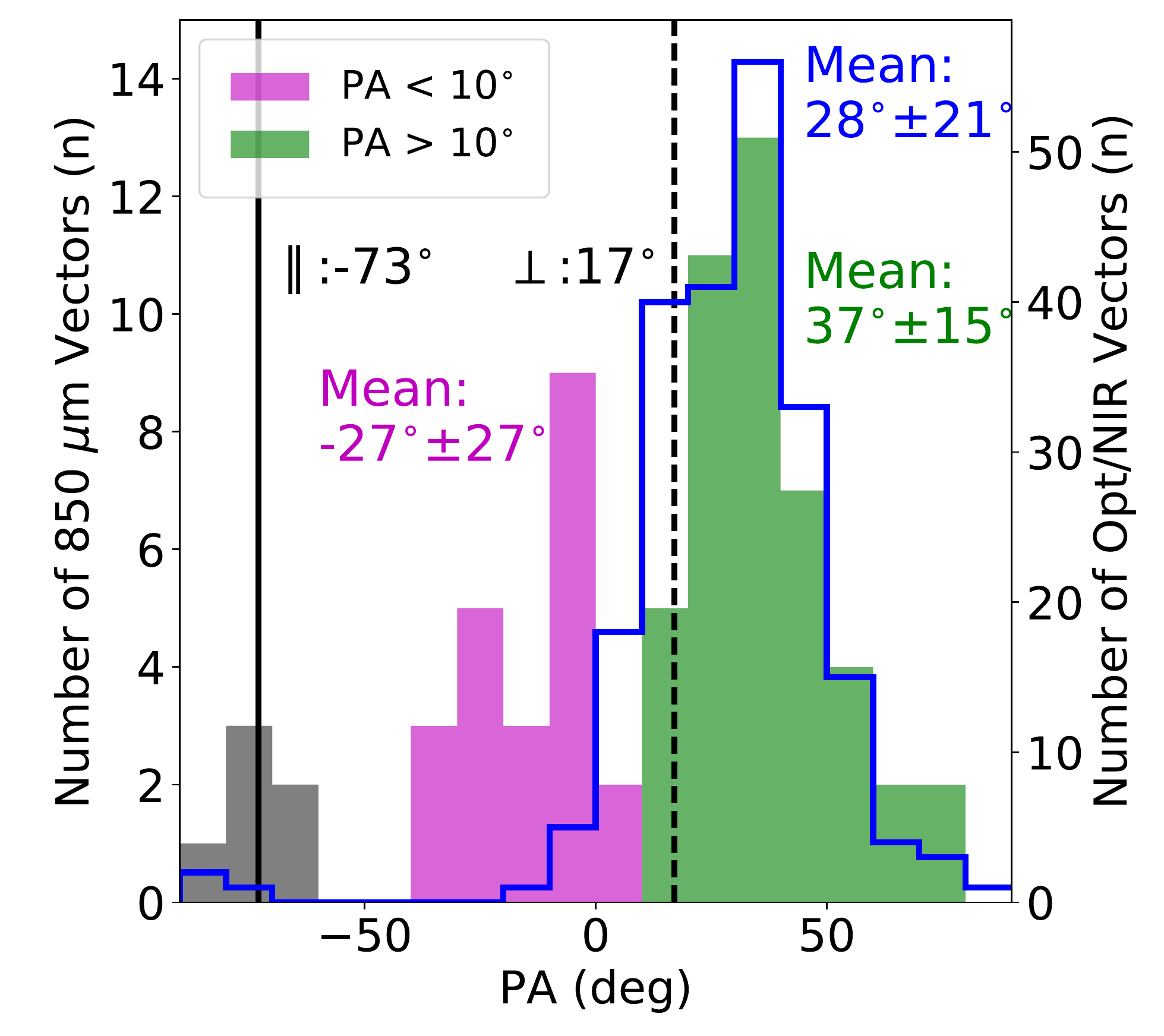}
\caption{Histogram of the magnetic field orientations. The colored histogram shows the 850 $\mu$m polarimetry data, and the blue histogram represents the optical/infrared data. The bin size of the histograms is set to 10\degr, similar to our typical uncertainties in $PA$. The $PA$=0\degr\ corresponds to north and $PA= +90$\degr\ is east. Our data show two major components with mean $PA$ of -27\degr\ (red) and 37\degr\ (green), separated by a dip at $\approx$ 10\degr, and a minor component peaked at -75\degr\ (gray). The black solid and dashed vertical lines label the orientation parallel and perpendicular to the large-scale filament, respectively, and the perpendicular orientation is consistent with the dip between the observed two components. }\label{fig:hist_2comp}
\end{figure}

\begin{figure}
\includegraphics[width=\columnwidth]{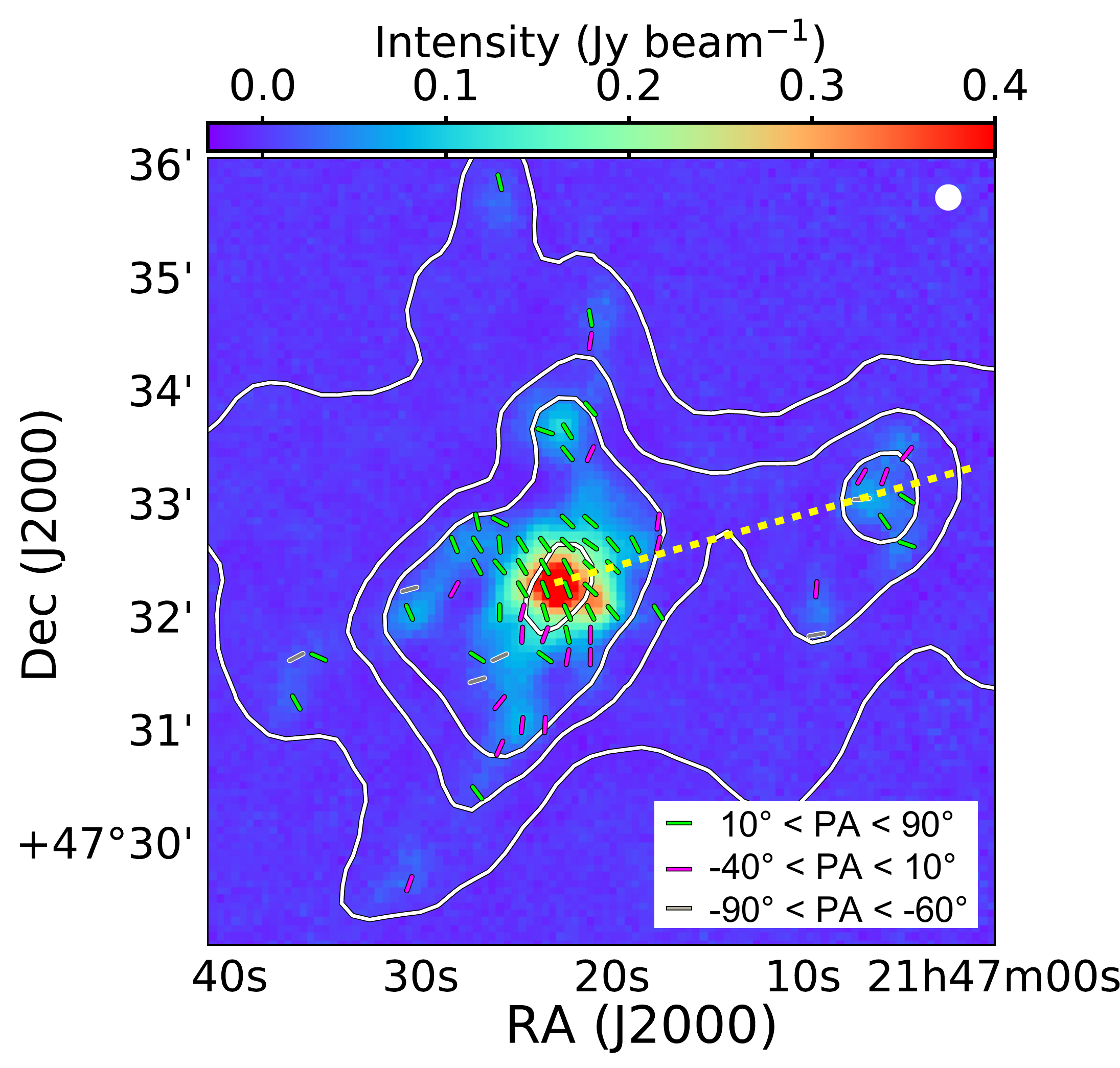}
\caption{The polarization vectors of the three components shown in \autoref{fig:hist_2comp}. The green, red, and gray vectors are associated with the components with $10\degr<PA<90\degr$, $-50\degr<PA<10\degr$, and $-90\degr<PA<-60\degr$, respectively. The white contours show the H$_2$ column densities of (0.5, 1, 1.5, and 3)$\times10^{21}$ cm$^{-2}$ \citep{ar11}, indicating the morphology of the large-scale filament. The yellow dashed line shows the direction across the intensity peak of the two clumps along the large-scale filament (PA=-73\degr), which we used to represent the major axis of the large-scale filament. The vectors from the two major components tend to distribute in the upper and bottom half of the system and are separated by the dashed line, which favors the possibility that the magnetic field is dragged by the large-scale main filament. The vectors from the minor component seem to be randomly distributed over the area, which are probably small-scale structures that we cannot resolve.}\label{fig:pmap_2comp}
\end{figure}

The Stokes $I$ map shows a central massive clump in which three filaments intersect. The observed morphology is consistent with the typical hub-filament structure. The central massive clump hosts $\sim 80\%$ of the total intensity within the system, and thus we recognize the central massive clump as the hub of the HFS. Three filaments are identified extending from the central hub to the north, east and south. The magnetic field revealed by our polarization map seems to have small angular dispersion, but also shows change of orientations from the north to the south.

To compare the magnetic fields in the observed HFS with the large-scale magnetic fields shown in \autoref{fig:field}, we plot a histogram of the position angles ($PA$s) of the local magnetic fields from POL-2 and WLE17 data within our field of view (diameter of 11\arcmin) in \autoref{fig:hist_2comp} with a bin size of 10\degr\ that is close to our mean $\delta PA$ of 8.5\degr.

The $PA$ histogram of our data shows two major components separated by a dip at 10\degr. The $PA > 10\degr$ component has a mean $PA$s of 37\degr\ and a $PA$ dispersion of 15\degr, which is similar to the large scale magnetic fields (28$\pm$21\degr). In contrast, the $PA < 10\degr$ component has a mean $PA$s of -27\degr\ and a $PA$ dispersion of 27\degr. The $PA$ difference of 64\degr\ between the two components is much greater than the $PA$ dispersion for large-scale magnetic fields and also our mean observational uncertainty (8.5\degr), suggesting that the observed magnetic field morphology is significantly different from the large-scale magnetic fields.

\autoref{fig:pmap_2comp} shows the locations of these two components. To represent the major axis of the main filament, we plotted the yellow dashed line that shows the direction across the intensity peaks of the two clumps along the parsec-scale filament. This major axis has a orientation of -73\degr, and roughly separates the spatial distribution of the two magnetic field orientation components. Within the HFS, the red and green components tend to distributed in the northern and southern half of the HFS; the tendency, however, is reversed in the western clump. In addition, the orientation perpendicular to the main filament (17\degr) is also close to the dip between the two components. These features favor the possibility that the magnetic fields in the HFS is curved along the main filament. In contrast, the WLE17 data only show a major peak (28\degr) in the $PA$ histogram, which is roughly perpendicular to the large-scale filament but with a $\approx 10\degr\ $ offset in $PA$. A minor component peaked at $\approx$ -75\degr\ is also shown by our data; however, this component is diffusely distributed over the area, and thus more vectors are needed to reveal these structures.

\subsection{Polarization Efficiency}\label{sec:peff}
\begin{figure}
 \includegraphics[width=\columnwidth]{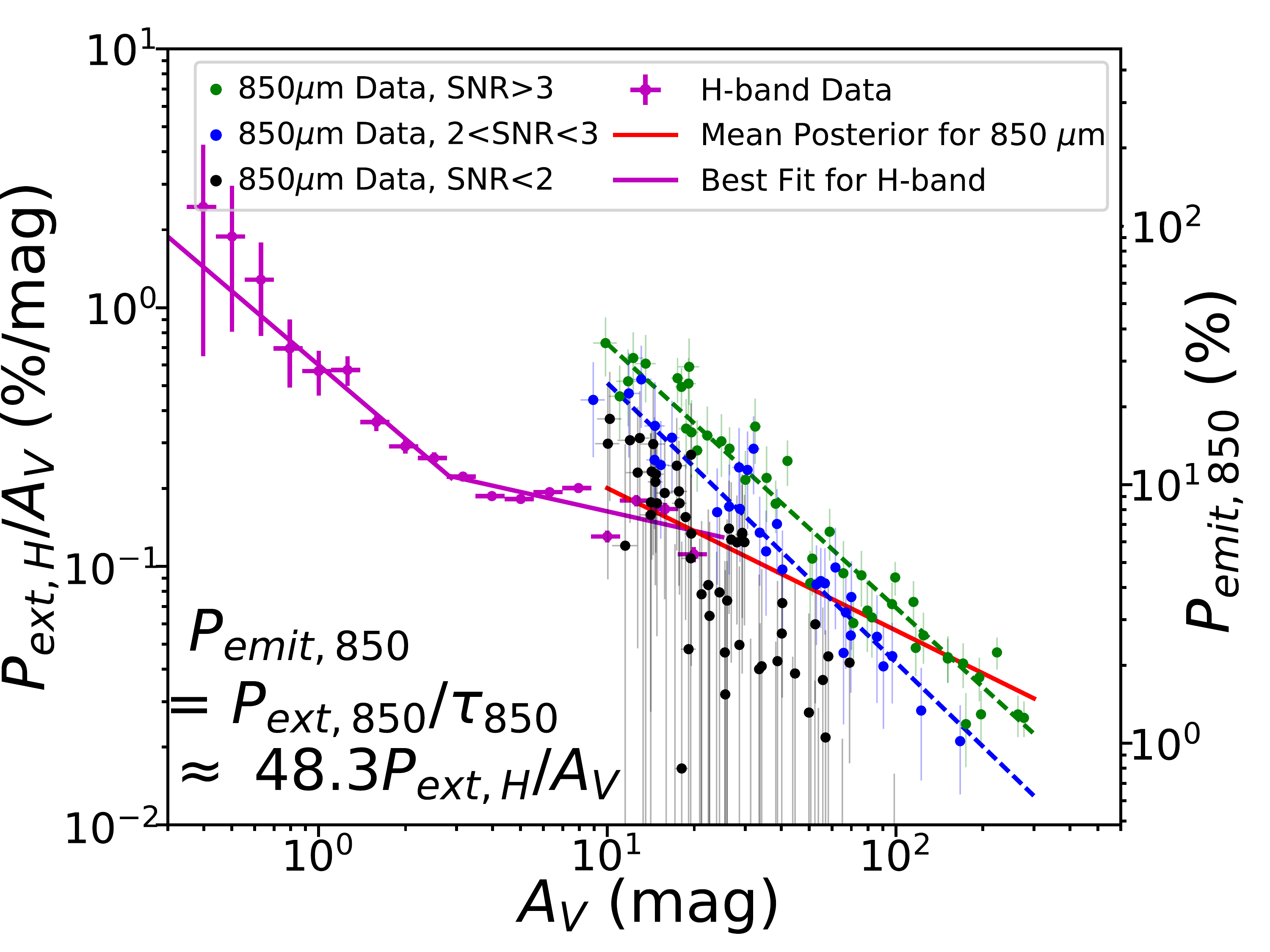}
\caption{Polarization efficiency versus $A_V$. The green, blue, and black points represent the POL-2 data with SNR greater than 3, between 2 and 3, and less than 2, respectively. The optical extinction of POL-2 data is derived from total intensities $I_{\mu}=\tau B_{\mu}(T)$ with temperatures derived in \citet{ar11} using the $Herschel$ data. The green and blue dashed lines show the best least-square fitting to the SNR $>3$ and $2<$ SNR $<3$ data with indices of -1.08 and -1.03, respectively.
The magenta points are the mean H-band polarization efficiency, observed across the whole IC5146 cloud system \citep{wa17}. The \pemit\ (for POL-2) and $PE$ (for H-band) values are shown in right and left y-axis, respectively. The magenta line represents the best fit for the H-band data \citep{wa17}, and the red line shows the prediction from the mean posterior for our data (\autoref{sec:pext}). These two fitting results are offset by a factor of 48.3 at $A_V$ = 20 mag, due to the wavelength-dependent optical depth of the aligned dust grains, which we use to scale and match the two data sets.
}\label{fig:peff}
\end{figure}

To investigate whether or not our polarization data trace the dust grains in high-extinction regions, we plot the 850 $\mu$m emission polarization fraction \pemit vs. $A_V$ in \autoref{fig:peff}. To reveal the complete \pemit--$A_V$ distribution, this figure includes all the data with $I/\delta I > 10$, and the data points are color-coded based on their SNR of \pemit.

To estimate the $A_V$, we calculated the $\tau_{850\mu m}$ from the observed $850 \mu m$ intensity using $I_{850\mu m}= \tau_{850\mu m} B(T_{dust})$, assuming that the dust emission is optically thin at $850 \mu m$. We used the dust temperature $B(T_{dust})$ derived in \citet{ar11} via fitting the \textit{Herschel} data at five wavelengths with a modified blackbody function assuming a dust emissivity index of 2. The dust temperature map and the Stokes $I$ map were both resampled on a 12\arcsec\ grid to match our polarization catalogue. The $\tau_{850\mu m}$ was converted to $A_V$ using the $R_V=3.1$ extinction curve in \citet{we01}.  We note that the extinction curve may vary at dense regions due to grain growth. If $R_V$ changes from 3.1 to 5.5 within the observed regions, we would underestimate the \pemit\ vs. $A_V$ slope by 10\%.

The \pemit\ are equivalent to the extinction polarization percentages divided by the optical depth (\pext/$\tau_{\lambda}$) in the optically thin case \citep{an15}, and so is proportional to the polarization efficiency ($PE$, defined as \pext/$A_V$). Thus, the observed \pemit\ vs. $A_V$ slope is equivalent to the $PE$ vs. $A_V$ slope.

We further plotted in \autoref{fig:peff} the $PE$ vs. $A_V$ revealed by WLE17 optical and infrared polarization data, to show how $PE$ varies in low $A_V$ regions. The $PE$ at 850 $\mu$m are in the form of $P_{ext,850}/\tau_{850}$, and the $PE$ obtained in H-band are represented by $P_{ext, H}/A_V$. Thus, a scaling factor $\frac{P_{ext,850}}{P_{ext, H}}\cdot\frac{A_V}{\tau_{850}}$ is required to convert the $PE$ at the two wavelengths, which is determined by the unknown dust properties \citep{an15}. Via matching the fitting results of $PE$ vs. $A_V$ relation in H-band (WLE17) and in 850 $\mu m$ bands (described in \autoref{sec:pext}) at $A_V$ = 20 mag, we found a scaling factor of 48.3, which we used to match the two data sets. This scaling factor is not a universal constant, as discussed by \citet{jo15}, and varies with physical conditions in different clouds.

%\section{Analysis}\label{sec:ana}
\subsection{Polarization Efficiency--$A_V$ Dependence}\label{sec:pext} 
To determine the \pemit\ vs. $A_V$ slope, the conventional approach is to apply a least-squares power-law fit to data selected by a SNR cut in \pemit. Following this approach, we fitted the $P/\delta P > 2$ and $> 3$ data with a power-law function. The best fit functions are shown in \autoref{fig:peff} by blue and green dashed lines, and the best fit power-law indices are $-1.02\pm0.02$ and $-1.08\pm0.02$, respectively. Nevertheless, since \autoref{fig:peff} shows that the \pemit--$A_V$ distribution is significantly truncated by the SNR cut, and also the best fit trends are very similar to the truncated boundary, it raises doubts on whether or not the fitting is biased by the sample selection.

We investigated how the sample selection biased on $P/\delta P$ affects the fitting to \pemit\ vs. $A_V$ distribution by performing Monte Carlo simulations of data sets with underlying \pemit\ vs. $A_V$ function and randomly generated measurement errors in \autoref{sec:appendix}. We found that the fitted power-index would be dominated by the SNR cut and approaches $-1$ rather than the true underlying value, if the \pemit\ vs. $A_V$ distribution is significantly truncated by the applied $P/\delta P$ selection criteria. Hence, to obtain an unbiased power-index, it is recommended to include the low $P/\delta P$ data, so that a complete probability density function (PDF) of \pemit\ can be recovered. Nevertheless, the use of low $P/\delta P$ data would break the Gaussian PDF assumption \citep{wa74,va06}, required for least-squares fit, and therefore we turn to use a Bayesian approach to fit the observed \pemit\ vs. $A_V$ distribution.

We used a Bayesian approach to apply a non-Gaussian PDF and fit the observed \pemit\ vs. $A_V$ trend with a power-law model. The Bayesian statistical framework provides a model fitting tool based on probability theory (see detailed introduction in \citealt{ja03}).
The general form of the Bayesian inference is
\begin{equation}
P(\theta|D)=\frac{P(\theta)P(D|\theta)}{P(D)}
\end{equation}
or
\begin{equation}
Posterior=\frac{Prior \times Likelihood}{Evidence},
\end{equation}
where D is the observed data and $\theta$ represents the model parameters. The posterior $P(\theta|D)$ describes the probability of the model parameters matching the given data, which is in what we are interested. The evidence $P(D)$ is the probability of obtaining the data, which mainly serves as a normalization factor for the posterior. The prior $P(\theta)$ serves as the initial guessed probability of the model parameters based on our prior knowledge. The likelihood $P(D|\theta)$ describes how likely it is for a given model parameter set to match the observed data. 

\begin{figure*}
\includegraphics[width=\textwidth]{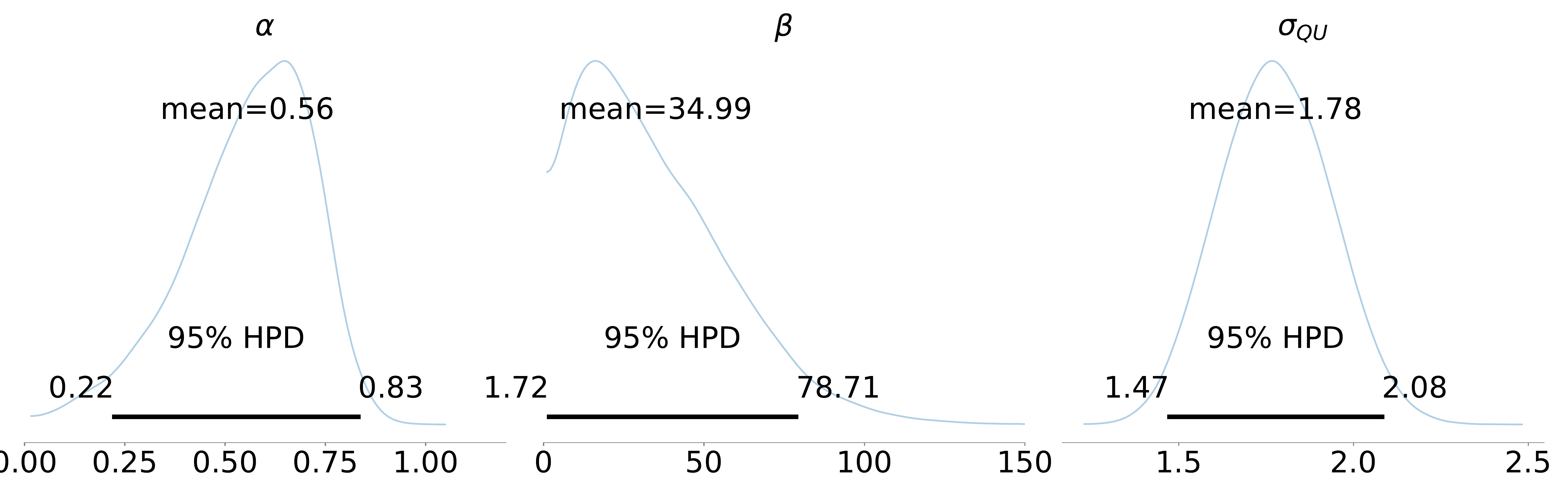}
\caption{PDF of the model parameters derived using Bayesian model fitting to our 12 arcsec data. The 95\% highest posterior density (HPD) intervals are plotted to represent the uncertainties. The derived $\alpha$ has a mean of 0.56 and a 95\% confidence interval from 0.22--0.83. The $\alpha$ value much lower than 1 suggests that the dust grains in $A_V\sim20$--300 mag can still be aligned with magnetic fields.
}\label{fig:baye_12asc}
\end{figure*}

\begin{figure}
\includegraphics[width=\linewidth]{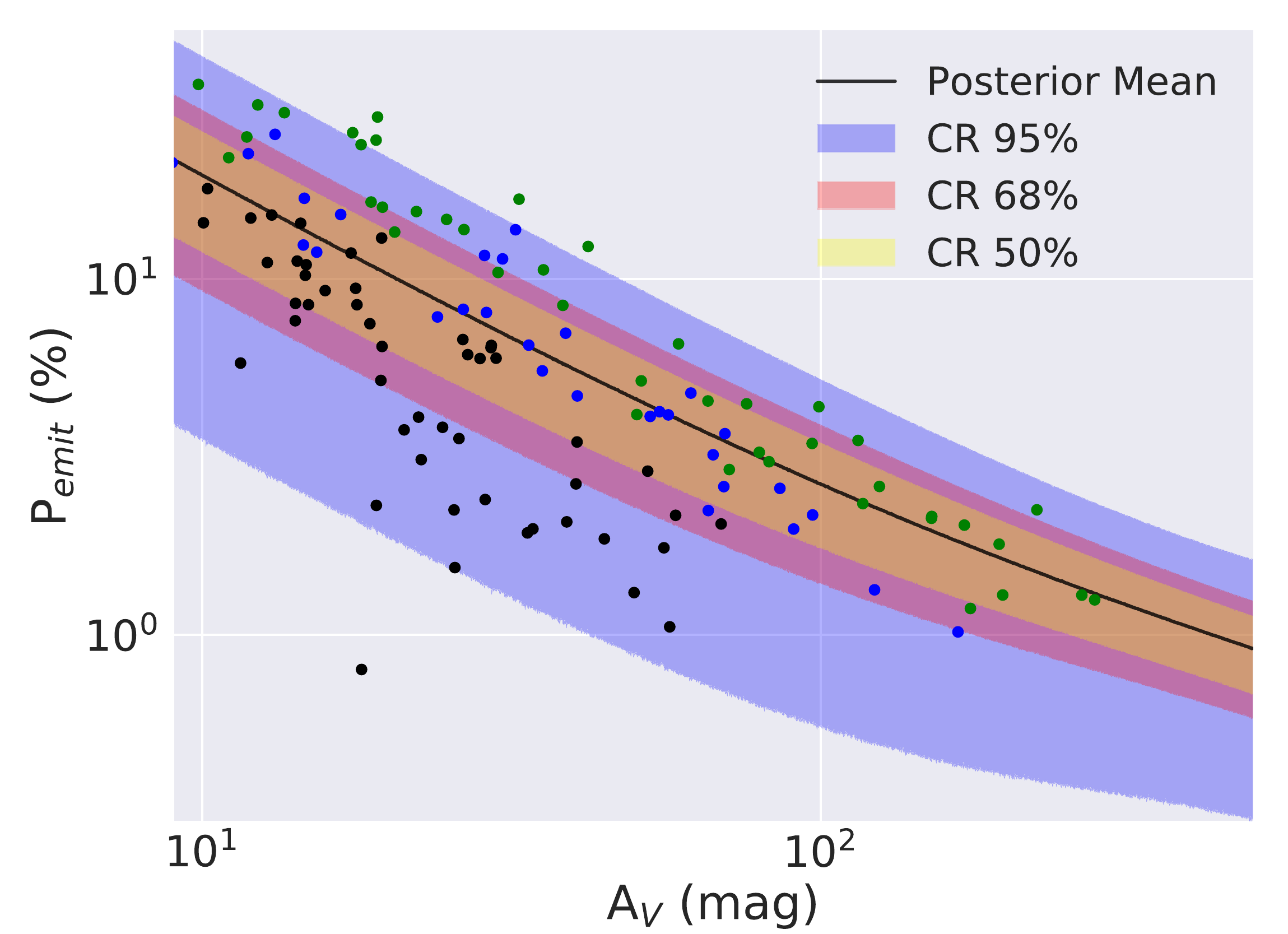}
\caption{The comparison between the Bayesian posterior prediction and the observations. The green, blue, and black points represent the POL-2 data with SNR greater than 3, between 2 and 3, and less than 2, respectively. The black line and colored regions show the mean, 95\%, 68\%, and 50\% confidence regions, predicted by the posterior shown in \autoref{fig:baye_12asc}, assuming a dust temperature of 13 K. Since the polarization error distribution is non-Gaussian and changes with $A_V$, the expected mean polarization is not simply a straight line on the logarithmic scale.}\label{fig:baye_fit}
\end{figure}

Assuming the measurements in Stokes $Q$ and $U$ have similar and Gaussian distributed noise, the probability density function of the observed polarization fraction has been known to follow the Rice distribution \citep{ri45,wa74,si85,qu12} 
\begin{equation}\label{eq:rice}
F(P|P_0)=\frac{P}{\sigma_P^{2}}exp\left[-\frac{P^2+P_0^2}{2{\sigma_P}^2}\right]I_0\left(\frac{PP_0}{\sigma_P^2}\right),
\end{equation}
where $P$ is the observed polarization fraction, $P_0$ is the real polarization fraction, $\sigma_P$ is the uncertainty in polarization fraction, and $I_0$ is the zeroth-order modified Bessel function. The likelihood function of polarization measurements is defined as:
\begin{equation}
L(P_0)=\displaystyle\prod_{i=1}^{n}F(P_n|P_0),
\end{equation}
where $P_n$ represents the nth measurement.

To perform the fit to the \pemit\ vs. $A_V$ trend using a Bayesian approach, we assumed the following power-law model such that 
\begin{equation}\label{eq:prior}
P_0 =\beta A_V^{-\alpha},
\end{equation}
where $\alpha$ and $\beta$ are the free model parameters, and $A_V$ is the observed visual extinction. The uncertainty in the polarization fraction is
\begin{equation}
\sigma_P=\sigma_{QU}/I.
\end{equation}
Here, the $I$ is the observed total intensity, and the $\sigma_{QU}$ is a free model parameter describing the dispersion in Stokes $Q$ and $U$, which has contributions both from the instrumental uncertainty and the intrinsic dispersion due to source properties. 
We then simply used uniform priors within reasonable limits as:
\begin{equation}
\begin{aligned}
 P(\alpha)& =
  \begin{cases}
   Uniform & 0< \alpha < 2 \\
    0 & otherwise \\
  \end{cases}\\
 P(\beta)& =
  \begin{cases}
    Uniform & 0 < \beta < 400 \\
    0 & otherwise \\
  \end{cases}\\
 P(\sigma_{QU})& =
  \begin{cases}
    Uniform & 0< \sigma_{QU} < 10 \\
    0 & otherwise. \\
  \end{cases}\\
\end{aligned}
\end{equation}
The Bayesian model fitting was performed with the Python Package PyMC3 \citep{sal16} via a Markov Chain Monte Carlo method using the Metropolis-Hastings sampling algorithm. The 12 arcsec pixel data were used for the fitting to ensure that each measurement is independent. 

The derived posterior of each model parameter is shown in \autoref{fig:baye_12asc}, and the 95\% highest posterior density (HPD) interval of each parameter is plotted to represent the uncertainty. The 95\%, 68\%, and 50\% confidence regions (CR) predicted by the posterior distribution in \pemit\ vs. $A_V$ space is shown in \autoref{fig:baye_fit}, assuming a dust temperature of 13 K. Since the error distribution of $P$ is asymmetric and also varies with $P/\sigma_P$ and $A_V$, the predicted $P$ vs. $A_V$ is not simply a straight line on a logarithmic scale, even though the input model is a power-law. Almost all the data points are well within the 95\% confidence regions predicted by the posterior. 

The derived $\alpha$ has a mean value of $0.56$ with a 95\% confidence interval from 0.22 to 0.83. The $\alpha$ derived by the Bayesian method is less steeper than the $\alpha\approx 1.0$ derived from the conventional approach, confirming that the conventional method is biased (see \autoref{fig:peff}). The $\alpha$ range of 0.22--0.83 includes the index of $0.25\pm0.06$ obtained from near-infrared polarization data in $A_V$ of 3--20 mag regions \citep[see][]{wa17}, and thus no significant difference in polarization efficiency was found between $A_V <$ 20 mag and $A_V =$ 20--300 mag regions (\autoref{fig:peff}). In addition, the fitted $\sigma_{QU}$ of 1.78 mJy~beam$^{-1}$ is greater than our estimated instrumental noise of 1.1 mJy~beam$^{-1}$, which indicates a significant intrinsic dispersion in polarization efficiency.

The value of $\alpha$ smaller than unity indicates that the extinction polarization fraction (\pext=$\tau$\pemit) increases with column density. Since the extinction polarization fraction is defined as tanh($\Delta\tau$) \citep{jo89}, where $\Delta\tau$ is the differential optical depth between two polarization directions, the increase of extinction polarization fraction indicates a higher amount of aligned dust grains. Hence, our results suggest that the dust grains in the IC5146 dense regions can still be aligned with magnetic fields. The $\alpha$ of $\sim 0.5$ is also predicted by the simulations based on radiative torque alignment theory \citep[e.g][]{wh08} in low density regions, where the radiation field is sufficiently strong to align dust grains.

\begin{figure*}
\includegraphics[width=\textwidth]{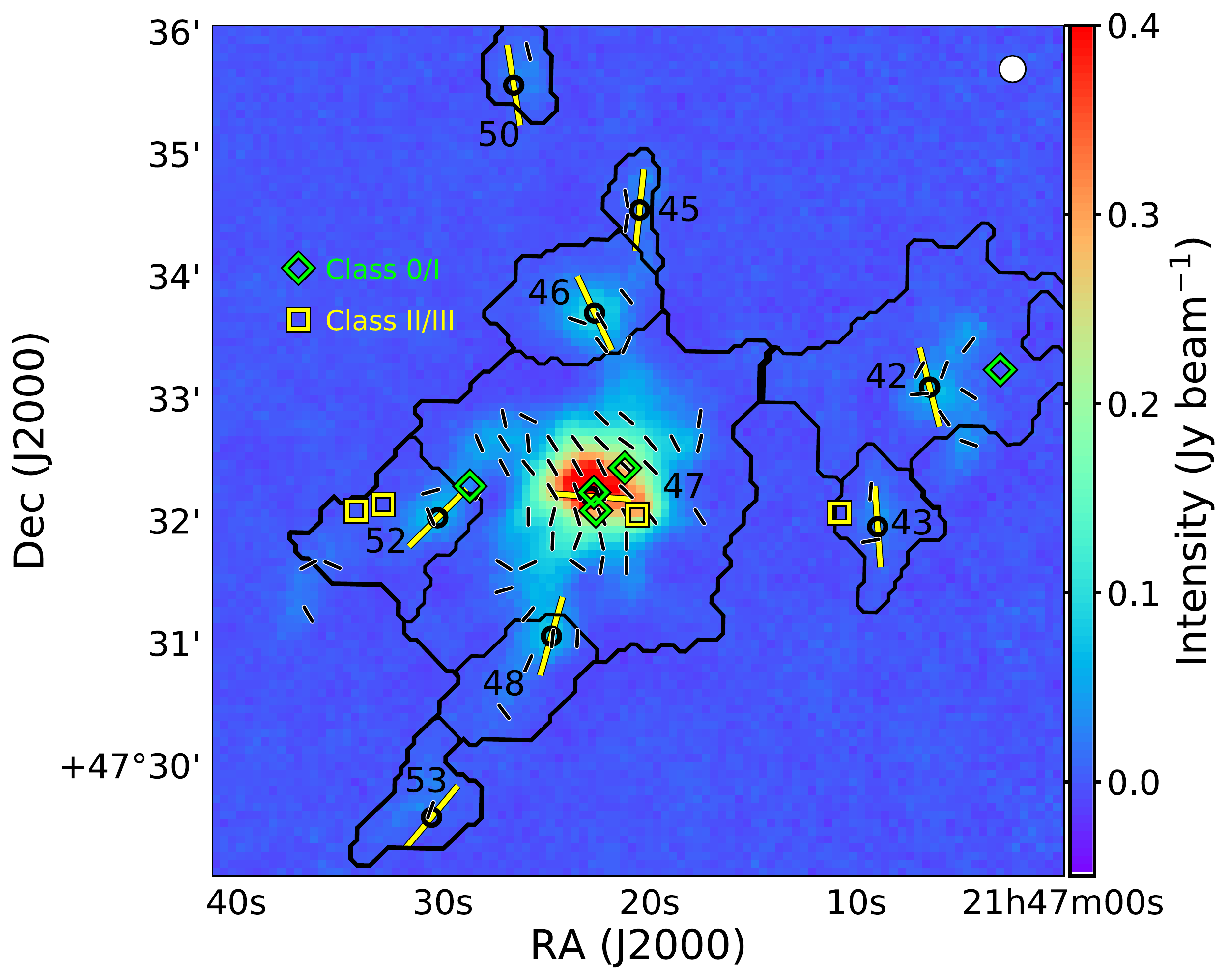}
\caption{The clumps identified by the JCMT Gould Belt survey \citep{jo17} overlaid on our polarization map. The image shows the 850 $\mu$m continuum emission. The black contours and circles represent the boundaries and the emission peaks of the identified clumps, respectively. The yellow vectors show the orientation of the major axis for each clump. The green diamonds and yellow boxes label the Class 0/I and II/III YSOs identified in \citet{du15}, respectively.}\label{fig:gbs}
\end{figure*}

Three possibilities could explain why the dust grains within these dense regions can still be aligned. First, since our target is an active star-forming region, the embedded young stars could be the sources of radiation needed to align the dust grains in dense regions. Second, WLE17 found that the dust grains in IC5146 have significantly grown from the diffuse ISM. These large dust grains could be aligned more efficiently by the radiation with longer wavelengths, which can penetrate the dense regions \citep{la07,ho09}. Third, the mechanical torques due to infalling gas and outflows in the star-forming regions could possibly align the dust grains in the absence of a radiation field \citep{la07b,ho18}. These possibilities will be further investigated in upcoming BISTRO papers probing the polarization efficiency in different environments.

\subsection{Orientation of Clumps and Magnetic Fields}\label{sec:clump}
To investigate whether or not magnetic fields influence the clump fragmentation within the IC5146 HFS, we examined the correlation between the magnetic field orientations and the clump morphologies. Based on the JCMT 850 $\mu$m Gould Belt Survey data, \citet{jo17} identified eight submillimeter clumps within the regions where we had polarization detections (\autoref{fig:gbs}). To represent the orientation of each clump, we used our total intensity map, and performed a 2D-Gaussian fit to each clump to find the position angles of its FWHM major axis. The 2D-Gaussian fit had typical orientation uncertainties of $\sim$ 10\degr. The obtained clump orientations are listed in \autoref{tab:clump} and plotted over the polarization map in \autoref{fig:gbs}.

To estimate the local magnetic field orientation within each clump, we calculated the mean magnetic field orientations by averaging the data within 3$\times$3 pixels at the clump intensity peaks. The size of 3$\times$3 pixels is comparable to the typical radius of these clumps ($\approx $20--40\arcsec, see \autoref{tab:clump}), and so the average represents the mean magnetic field orientation over the dense center of these clumps. In addition, The 3$\times$3 pixels also provide an estimation of orientation dispersion, which were used as the uncertainties of the averaged orientations; if only one vector was obtained for a given clump, the instrumental uncertainty would be used. To handle the $\pm$180\degr\ ambiguity, the mean and dispersion of the magnetic field $PA$s were calculated in a new coordinate system, where the $PA$ dispersion were minimized, and the calculated results were converted back to the standard coordinate system.

\begin{figure}
\includegraphics[width=\columnwidth]{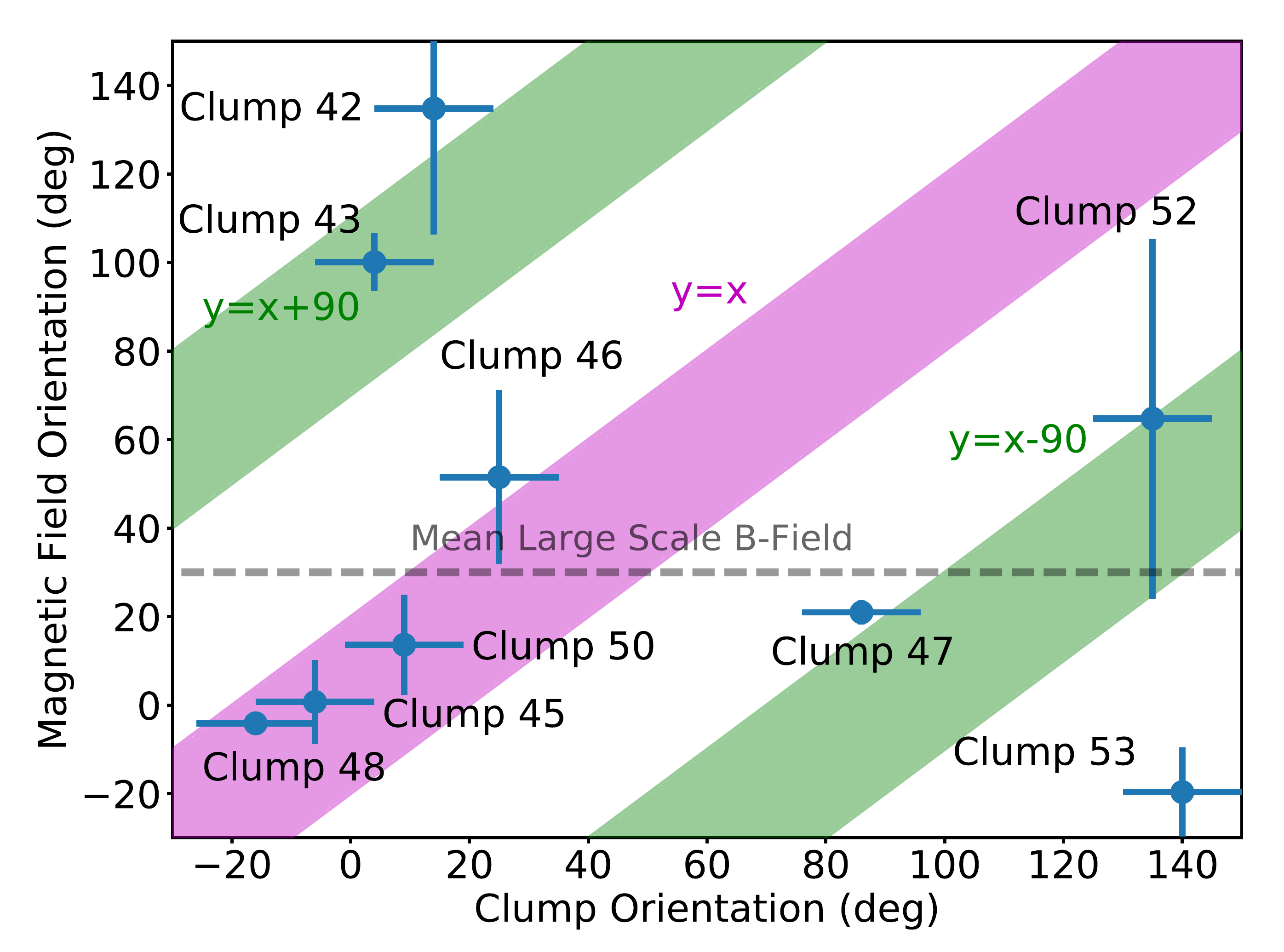}
\caption{The comparison of orientations between clumps and magnetic fields. The clump orientations are obtained from 2D-Gaussian fits to the CO-subtracted intensity, and the magnetic field orientations are averaged from the polarization detection within the $3\times3$ pixels at the clump intensity peak. The gray dashed line labels the mean $PA$ of the large-scale magnetic field from WLE17 data. The magenta region represents where the orientation of clumps and magnetic fields are equal, and the green regions show where the orientation of clumps and magnetic fields are perpendicular. All our clumps are close to either magenta or green regions. }\label{fig:papa}
\end{figure}

The derived local magnetic field orientations versus the clump orientations are plotted in \autoref{fig:papa}. The comparison between the clump axis and the magnetic field orientations in the clump is limited by the small number of statistics. Nevertheless, there appears to be a tendency that the observed clumps are likely either parallel or perpendicular to the mean magnetic field orientation within $\sim$20\degr, as shown in \autoref{fig:papa}. The upcoming BISTRO data would provide a much bigger sample set from various systems to statistically confirm this tendency. If the tendency is real, it would suggest that the magnetic fields are a key factor in the fragmentation of these clumps. We note that clumps 43 and 52 only contain two polarization vectors which are almost perpendicular to each other, and thus the mean magnetic field orientations are not meaningful for these two cases. Since the orientation of these two clumps are still parallel to one of the vectors and perpendicular to the other, these two clumps, however, are still consistent with the tendency.

\begin{deluxetable*}{ccccccccccc}
\tablecaption{Geometric and Polarization Properties of the Clumps}
%\tablenum{1}
\tablehead{\colhead{ID\tablenotemark{a}} & \colhead{Total Mass\tablenotemark{a}} & \colhead{Major Axis} & \colhead{Minor Axis} & \colhead{R$_{eff}$} & \colhead{Clump Orientation} & \colhead{B-field PA$_{peak}$\tablenotemark{b}} & \colhead{$\sigma_{NT}$} & \colhead{$\alpha_{vir}=\frac{\textrm{M$_{vir}$}}{M_{clump}}$\tablenotemark{c}} & \colhead{$\alpha_{vir,B}$} \\ 
\colhead{} & \colhead{(M$_{\odot}$)} & \colhead{(arcsec)} & \colhead{(arcsec)} & \colhead{(arcsec)} & \colhead{(deg)} & \colhead{(deg)} & \colhead{{(km~s$^{-1}$)}} & \colhead{} & \colhead{} & \colhead{}} 
\startdata
42 & 11$\pm3$ & 51$\pm2$ & 33$\pm2$ & 43$\pm2$ & $14\pm10$ & 134.8$\pm$28.5 & $0.25\pm0.01$ & 0.9$\pm0.2$ & ...\\
43 & 2.0$\pm0.5$ & 29$\pm1$ & 16$\pm1$ & 22$\pm1$ & 4$\pm10$ & 100.1$\pm$6.6 & $0.12\pm0.01$ & 1.4$\pm0.3$ & ...\\
45 & 0.77$\pm0.18$ & 37$\pm3$ & 16$\pm1$ & 24$\pm2$ & 174$\pm10$ & 0.7$\pm$9.5 & $0.18\pm0.02$ & 5.1$\pm1.0$ & ...\\
46 & 6.4$\pm1.6$ & 26$\pm2$ & 23$\pm1$ & 24$\pm2$ & 25$\pm10$ & 51.5$\pm$19.7 & $0.21\pm0.03$ & 0.7$\pm0.1$ & ...\\
47 & 85$\pm20$ & 40$\pm2$ & 36$\pm2$ & 38$\pm2$ & 86$\pm10$ & 21.0$\pm$2.7 & $0.36\pm0.12$ & 0.2$\pm0.1$ & 0.2--1.0\tablenotemark{d}\\
48 & 6.0$\pm1.5$ & 35$\pm3$ & 25$\pm1$ & 30$\pm3$ & 164$\pm10$ & -4.1$\pm$1.6 & $0.14\pm0.01$ & 0.7$\pm0.1$ & ...\\
50 & 0.97$\pm0.24$ & 29$\pm1$ & 18$\pm1$ & 23$\pm1$ & 9$\pm10$ & 13.6$\pm$11.4 & ... & ... & ...\\
52 & 7.6$\pm1.9$ & 31$\pm2$ & 17$\pm1$ & 23$\pm2$ & 135$\pm10$ & 64.7$\pm$40.7 & $0.29\pm0.01$ & 0.9$\pm0.2$ & ...\\
53 & 1.7$\pm0.4$ & 34$\pm3$ & 18$\pm1$ & 24$\pm3$ & 140$\pm10$ & -19.7$\pm$10.1 & ... & ... & ...\\
\enddata
\tablenotetext{a}{The clumps ID and mass were obtained from \citet{jo17} but the masses were scaled to a distance of $813\pm106$ pc.}
\tablenotetext{b}{Mean magnetic field orientation averaged using the $3\times3$ pixels at the intensity peaks.}
\tablenotetext{c}{The virial masses of clumps were calculated considering the support from thermal pressure and turbulence.}
\tablenotetext{d}{If the inclination angle of the magnetic field with respect to the line of sight is greater than 15\degr.}
\label{tab:clump}
\end{deluxetable*}
%\FloatBarrier

We further plot in \autoref{fig:papa} the mean large-scale magnetic field orientation, 28\degr, obtained from WLE17. Only Clump 47 has a magnetic field orientation similar to the large-scale magnetic fields within 20\degr. All other clumps are aligned either parallel or perpendicular with the local magnetic field and show no significant correlation with the large-scale magnetic fields. Hence, these clumps are more likely formed after the local magnetic fields were distorted by the process of the clump formation. 

\subsection{Magnetic Field Strength in IC5146}\label{sec:str}
The Davis-Chandrasekhar-Fermi (DCF) method \citep{da51,ch53} is commonly used to evaluate the magnetic field strength from dust polarizations. The DCF method assumes that turbulent kinematic energy and the turbulent magnetic energy are in equipartition, and hence the magnetic field strength can be estimated using
\begin{equation}\label{eq:CF}
B_{pos}=Q~\sqrt[]{4\pi \rho}\frac{\sigma_v}{\delta \phi_{intrinsic}},
\end{equation}
where $\delta \phi$ is the intrinsic angular dispersion of the magnetic fields, $\sigma_v$ is the velocity dispersion, $\rho$ is the gas density, and Q is a factor accounting for the complicated magnetic field and inhomogeneous density structure. \citet{os01} suggested that $Q=0.5$ yields a good estimation of magnetic field strength on the plane of sky if magnetic field angular dispersion is less than 25\degr .

\subsubsection{Magnetic Field Angular Dispersion}\label{sec:ad}
We used the 12\arcsec\ pixel polarization data to calculate the magnetic field angular dispersion to ensure that all vectors we used are independent measurements. To avoid small number statistics (less than 10 vectors), we only perform the angular dispersion estimation using the polarization vectors (45 vectors) within the central hub (Clump 47) (\autoref{fig:gbs}). 

The DCF method requires an estimation of magnetic field distortion caused by turbulence, and the underlying magnetic field geometry might bias the estimation. Thus, we calculated the magnetic field angular dispersion in a local area to avoid the angular dispersion contribution from the large-scale nonuniform magnetic field geometry. Specifically, we selected each 24\arcsec$\times$24\arcsec\ box (i.e. the width of 2 independent beams), and calculated the mean and the corresponding sum of squared differences (SSD=$\sum_{i=1}^{n}(\bar{PA}-PA_i)^2$) of the $PA$ using the, at most, 9 vectors within each box. The SSD from all boxes were averaged with inverse-variance weighting, and the square root of the mean SSD was taken as the observed angular dispersion. Next, the mean instrumental uncertainties ($\delta \phi_{ins}$) of $8.0\degr$ were removed from the observed angular dispersion ($\delta \phi_{obs}$) to obtain the intrinsic dispersion ($\delta \phi_{intrinsic}$) via 
\begin{equation}
\delta \phi^2_{intrinsic}=\delta \phi^2_{obs}-\delta \phi^2_{ins}.
\end{equation}
With these corrections, the calculated $\delta \phi_{intrinsic}$ for clump 47 is $17.4\degr\pm0.6\degr$.

\subsubsection{Velocity Dispersion}
To estimate the velocity dispersion, we used the C$^{18}$O (J = 3-2) spectral data taken by \citet{gr08} with the JCMT HARP receiver \citep{bu09}. CO and its isotopomers are well mixed with H$_2$ and are commonly used to trace the gas kinematics. The C$^{18}$O (J = 3-2) line, in particular, is expected to trace gas with volume density up to $\sim 10^5$ cm$^{-3}$ \citep[e.g.,][]{di07}, which is comparable to the densities of our target field. In addition, the C$^{18}$O (J = 3-2) line is likely optically thin in this field \citep{gr08}, and so it traces the kinematics of all the gas in the clump. Therefore, we assumed that the C$^{18}$O (J = 3-2) line width can well represent the gas velocity dispersion in our observing regions.

The C$^{18}$O data reveals, at least, three velocity components within the central hub, peaked at 3.7, 4.1, and 4.5 km~s$^{-1}$. Because the three components have very similar velocities, and also multiple YSOs have been identified in the central hub by \citet{ha08}, the multiple components are possibly structures within the hub, instead of foreground or background components. We performed a multi-component Gaussian fit to estimate the C$^{18}$O (J = 3-2) line width, using the python package PySpecKit \citep{gi11}. We only accepted the fitted Gaussian components with amplitudes larger than 5 $\sigma$. To estimate the thermal velocity dispersion, we adopted a gas kinematic temperature ($T_\mathrm{kin}$) of 10$\pm 1$ K which is the same as the excitation temperature estimated from $^{13}$CO (J = 3-2) line in \citet{gr08}, leading to $\sqrt{\frac{k_B T_\mathrm{kin}}{m_{\mathrm{C}^{18}\mathrm{O}}}}=0.05\pm0.01$ km~s$^{-1}$. The thermal velocity dispersions were then removed from the fitted line widths to obtain the non-thermal velocity dispersions via
\begin{equation}\label{eq:vdisp}
\sigma^2_{NT}=\sigma^2_{obs} - \frac{k_B T_\mathrm{kin}}{m_{\mathrm{C}^{18}\mathrm{O}}}
\end{equation}
where $\sigma_{NT}$ is the non-thermal velocity dispersion, $\sigma_{obs}$ is the observed C$^{18}$O Gaussian line width, and $m_{\mathrm{C}^{18}\mathrm{O}}$ is the molecular weight. The inverse-variance weighted mean of the $\sigma_{obs}$ and $\sigma_{NT}$ of all velocity components over the central hub was 0.37 and 0.36 km~s$^{-1}$, respectively, and the dispersion of $\sigma_{obs}$ of 0.12 km~s$^{-1}$ among the central hub was used as its uncertainty. 

\subsubsection{Volume Density}
\citet{jo17} estimated the total mass of the clumps within the IC5146 cloud using JCMT 850 $\mu$m data assuming a distance of 950 pc. The derived total masses of the central hub were scaled down to a distance of $813\pm106$ pc and are listed in \autoref{tab:clump}. We assume that the thickness of the hub is equal to the geometric mean of the observed major and minor axis, obtained from 2D-Gaussian fit listed in \autoref{tab:clump}, and the uncertainty of the thickness is assumed to be the difference between the observed major and minor axis. The mean volume density of the hub is then estimated using its total mass and ellipsoid volume. The calculated H$_2$ volume densities ($n_{H_2}$) for Clump 47 are 9.8$\pm2.4\times10^5$ cm$^{-3}$. We note that our estimated radius is underestimated due to the unknown inclination angle ($i$) of the clump, and thus the volume densities we estimated here are only upper limits.

\begin{deluxetable}{cccccc}
%\centering
\tablecaption{Derived magnetic field Strength of the Clump 47\label{tab:CF}}
\renewcommand{\thetable}{\arabic{table}}
\tablenum{2}
\tablehead{\colhead{ID} & \colhead{$\sigma_{NT}$} & \colhead{$\delta \phi$} & \colhead{$n_{H_2}$} & \colhead{$B_{pos}$} & \colhead{$\lambda$} \\
\colhead{} & \colhead{(km~s$^{-1}$)} & \colhead{(deg)} & \colhead{(cm${^-3}$)} & \colhead{(mG)} & \colhead{}}
\startdata
47 & $0.36\pm0.12$ & $17.4\pm0.6$ & $(9.8\pm2.4) \times 10^5$ & $0.5\pm0.2$ & $1.3\pm0.4$ \\
\enddata
\tablecomments{The magnetic field strength estimated using DCF method. $\sigma_{NT}$, $\delta \phi$, $n_{H_2}$, $B_{pos}$, and $\lambda$ represent the velocity dispersion, magnetic field angular dispersion, H$_2$ volume density, plane of sky magnetic field strength, and mass-to-flux ratio, respectively. }
{\addtocounter{table}{-1}}
\end{deluxetable}

\subsubsection{Magnetic Field Strength and Mass-to-Flux Ratio}\label{sec:str_m2f}
Using \autoref{eq:CF} and the quantities estimated above (\autoref{tab:CF}), the plane-of-sky magnetic field strength ($B_{pos}$) is estimated to be $0.5\pm0.2$ mG.
To evaluate the relative importance of magnetic fields and gravity in the central hub, we calculate the mass-to-flux critical ratio via
\begin{equation}
\lambda_{obs}=\frac{(M/\Phi)_{obs}}{(M/\Phi)_{cri}},
\end{equation}
where the observed mass-to-flux ratio is
\begin{equation}
(M/\Phi)_{obs}=\frac{\mu m_{H}N_{H_2}}{B_{pos}},
\end{equation}
where $\mu$=2.8 is the mean molecular weight per H$_2$ molecule, and the $(M/\Phi)_{cri}$ is the critical mass-to-flux ratio defined as
\begin{equation}
(M/\Phi)_{cri}=\frac{1}{2\pi \sqrt{G}}
\end{equation}
\citep{na78}. Due to the unknown inclination of the clumps, the observed mass-to-flux ratio $\lambda_{obs}$ is also only an upper limit. \citet{cr04} suggested that a statistically average factor of $\frac{1}{3}$ could be used to estimate the real mass-to-flux ratio accounting for the random inclinations for an oblate spheroid core, flattened perpendicular to the orientation of the magnetic field. Since we have shown that the central clump is elongated with its major axis perpendicular to the local magnetic field, we adopt a factor of $\frac{1}{3}$ to estimate the mass-to-flux ratio via
\begin{equation}
\lambda=\frac{1}{3} \lambda_{obs}.
\end{equation}

The estimated mass-to-flux ratio for the central clump is $1.3\pm0.4$. The DCF method often tends to overestimate the magnetic field strength, since the effect of integration over the telescope beam and along the line of sight might smooth out part of the magnetic field structure, resulting in an underestimated angular dispersion \citep{he01,os01,cr12}. In addition, our target region has a complicated velocity structure, and therefore the measured velocity dispersion might have contributions from gas accretion or contraction motions instead of only isotropic turbulence, also leading to an overestimated magnetic strength. Hence, our estimation of the mass-to-flux ratio only represents a lower limit. A mass-to-flux ratio $\ga 1.0$ suggests that the central hub is super-critical, and that magnetic fields and gravity are comparably important at sub-parcsec scale.

\subsubsection{Angular Dispersion Function}\label{sec:SF}
\citet{hi09} developed an alternative method to improve the DCF method using an polarization angular dispersion function to accurately extract the turbulent component from the polarization data. \citet{hou09} further generalized the angular dispersion function method (hereafter ADF method) by including the effect of signal integration along the thickness of the clouds and over the telescope beam. In this section, we test whether or not the magnetic field strength estimated using the ADF method is significantly different from our estimation in \autoref{sec:str_m2f}.

The ADF method assumes that the magnetic fields in clouds are combinations of ordered large-scale component $B_0$ and turbulent component $B_t$, and the ratio of these two components determines the intrinsic polarization angular dispersion that
\begin{equation}
\delta \phi_{intrinsic} = \left[\frac{\langle B_t^2\rangle}{\langle B_0^2\rangle}\right]^{\frac{1}{2}},
\end{equation}
where $\langle...\rangle$ denotes an average. Hence, the DCF equation (\autoref{eq:CF}) can be written as 
\begin{equation}
B_{pos}=\sqrt[]{4\pi \rho}~\sigma_v \left[\frac{\langle B_t^2\rangle}{\langle B_0^2\rangle}\right]^{-\frac{1}{2}}.
\end{equation}

The detailed derivation given by \citet{hi09} and \citet{hou09} shows that the ratio of turbulent to magnetic energy can be estimated from the angular dispersion function using the following equation:
\begin{equation}\label{eq:SF}
1-\langle\cos[\Delta\Phi(\ell)]\rangle \simeq \frac{1}{N}\frac{\langle B_t^2\rangle}{\langle B_0^2\rangle}(1-e^{-\ell^2/2(\delta^2+2W^2)})+a\ell^2,
\end{equation}
where $\Delta \Phi (\ell)$ is the difference in the polarization angle measured at two positions separated by a distance $\ell$. The quantities $\delta$ and $a$ are unknown parameters, representing the turbulent correlation length and the first order Taylor expansion of the large-scale magnetic field structure. The quantity $W$ is the telescope beam radius, which is 7.3 arcsec at 850 $\mu$m. The quantity $N$ is the number of turbulent cell observed along the line of sight and within the telescope beam, and can be estimated from:
\begin{equation}\label{eq:ncell}
N=\Delta'\frac{\delta^2+2W^2}{\sqrt{2\pi}\delta^3},
\end{equation}
where $\Delta'$ is the effective cloud thickness, which is assumed to be the clump effective radius. Via fitting the above equation to the observed $1-\langle\cos[\Delta\Phi(\ell)]\rangle$ vs. $\ell$ distribution, the three unknown parameters $\delta$, $\frac{\langle B_t^2\rangle}{\langle B_0^2\rangle}$, and $a$ can be derived

\begin{figure}
\includegraphics[width=\columnwidth]{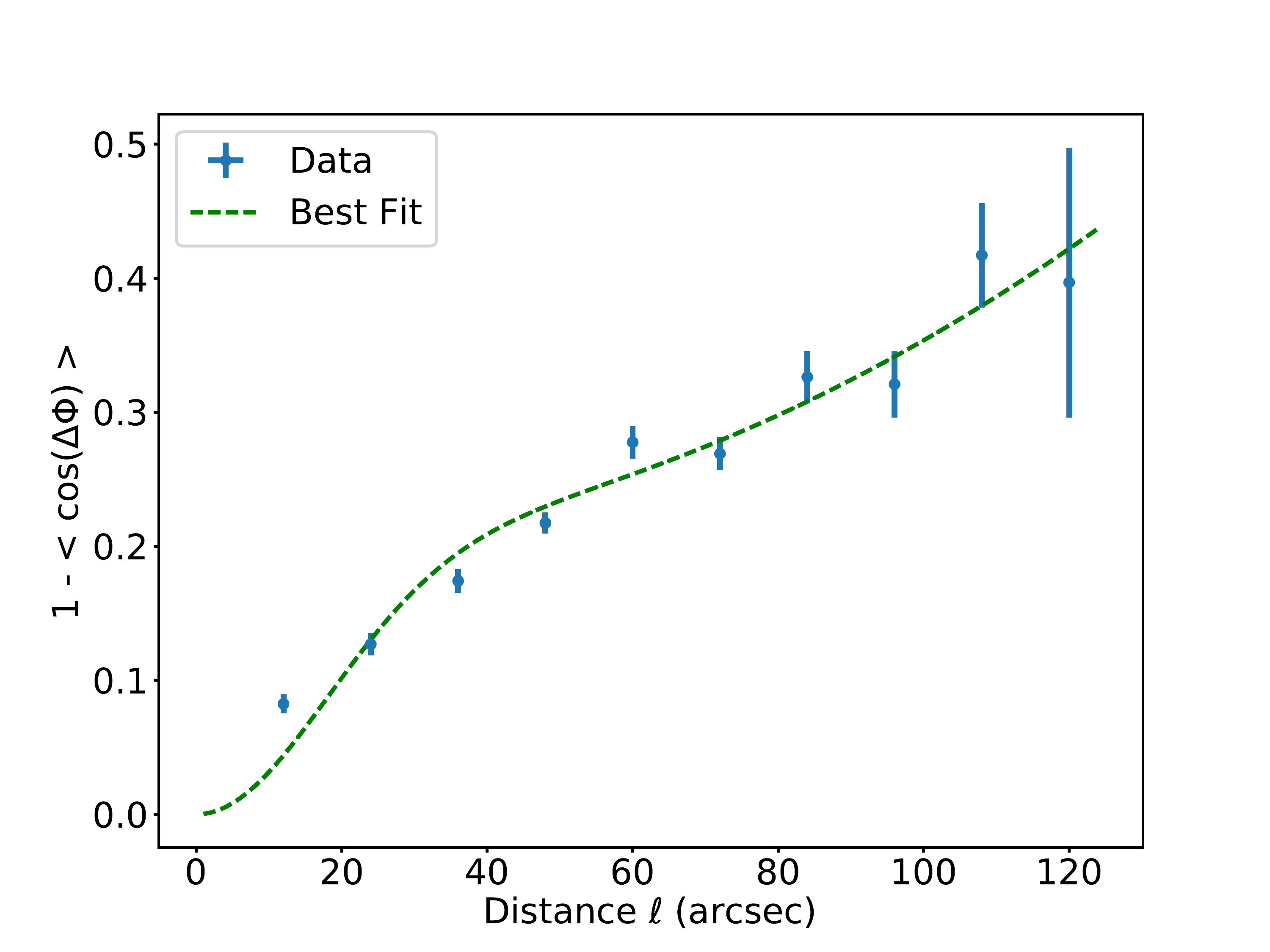}
\caption{The angular dispersion function $1-\langle\cos[\Delta\Phi(\ell)]\rangle$ as a function of the distance $\ell$. The mean $\Delta\Phi(\ell)$ was calculated in bins of 12 arcsec. The green dashed line shows the best fit of \autoref{eq:SF} to the data.}\label{fig:SF}
\end{figure}

We applied the ADF method to estimate the magnetic field strength in Clump 47 using the same selected polarization vectors as in \autoref{sec:ad}. We calculated the $\cos[\Delta\Phi(\ell)]$ and $\ell$ from each pair of the polarization vectors within the Clump 47, and the results are averaged in bins of width $\ell$ = 12 arcsec to estimate the angular dispersion function $1-\langle\cos[\Delta\Phi(\ell)]\rangle$ vs. $\ell$. The calculated angular dispersion function is plotted in \autoref{fig:SF}. We fitted the observed angular dispersion function using \autoref{eq:SF} and \autoref{eq:ncell}, and the best fitting parameters are shown in \autoref{tab:SF}. The obtained turbulent to magnetic energy ratio $\frac{\langle B_t^2\rangle}{\langle B_0^2\rangle}$ is $0.33\pm0.04$, suggesting that the turbulent magnetic field component is weaker than the ordered large-scale magnetic field. With the previously derived gas velocity and the volume density (\autoref{tab:CF}), the magnetic field strength is estimated to be $0.5\pm0.2$ mG, which is consistent with our estimation using the DCF method ($0.5\pm0.2$ mG) within the uncertainties.

\begin{table}[]
%\centering

\caption{Magnetic field Strength of Clump 47 using the ADF method\label{tab:SF}}
\renewcommand{\thetable}{\arabic{table}}
\tablenum{3}
\begin{tabular}{ccccc}
\hline
\hline
\multicolumn{3}{c}{Fit Result} & \multicolumn{2}{c}{Derived Quantities} \\
\cmidrule(lr){1-3} \cmidrule(lr){4-5}
$\delta$ & $\frac{\langle B_t^2\rangle}{\langle B_0^2\rangle}$ & $a$ & $N$ & $B_{pos}$ \\
(arcsec) &  & (arcsec$^{-2}$) & & (mG) \\
 \hline
$14.1\pm4.2$ & $0.33\pm0.04$ & $(1.5\pm0.3)\times10^{-5}$ & 1.7$\pm$0.5 & $0.5\pm0.2$ \\
\hline
\hline
\end{tabular}
%\tablecomments{The magnetic field strength estimated using the ADF method assuming a distance of 460 pc and 950 pc.}
\end{table}

\subsubsection{Alfv\'{e}nic Mach Number}\label{sec:Ma}
The turbulent Alfv\'{e}nic Mach number ($M_A$) describes the relative importance of magnetic fields and turbulence, and hence it is a key factor in most of cloud evolution models \citep[e.g.,][]{pa01,na08}. In the sub-Alfv\'{e}nic case ($M_A \le$ 1), magnetic fields are strong enough to regulate turbulence, and cause an organized magnetic field and cloud structure. In the super-Alfv\'{e}nic case ($M_A >$ 1), the turbulence can efficiently change the magnetic field morphology, and the magnetic field morphology is expected to be random.

The Alfv\'{e}nic Mach number can be estimated from the angular dispersion of the magnetic field if the same assumptions as for the DCF method are made. In doing so, the definition of the Alfv\'{e}nic Mach number 
\begin{equation}
M_A= \frac{\sigma_{NT}}{V_A}= \frac{\sigma_{NT}\sqrt{4\pi \rho}}{B}
\end{equation}
can be combined with the equation of the DCF method (\autoref{eq:CF}), yielding
\begin{equation}
M_A= \frac{\delta \phi \cdot \sin\theta}{Q},
\end{equation}
where $\theta$ is the inclination of the magnetic fields, with respect to the line of sight, so that $B_{pos}=B \sin\theta$. For $Q=0.5$, the obtained magnetic field angular dispersion 17\degr\- corresponds to $M_A$ of $0.6\sin\theta$, and hence the central hub is likely sub-Alfv\'{e}nic.

\subsection{Gravitational Stability of Clumps}
In this section, we use the virial analysis to investigate whether or not thermal pressure, turbulence, and magnetic fields are sufficient to support clumps against gravity. If a clump with uniform density is supported by only thermal pressure and turbulence, the virial mass (M$_{vir}$) is 
\begin{equation}
M_{vir}=\frac{5R_{eff}}{G}(\sigma^2_{NT}+c_s^2)
\end{equation}
\citep{be92,pi11,liu18}, where $R_{eff}$ is the geometric mean of major and minor radius, and $c_s=0.19$ km~s$^{-1}$ is the sound speed for a kinematic temperature of 10 K and mean molecular weight. Virial mass is the maximum mass of a stable clump with the support from kinematic and thermal energy. Hence, a clump mass greater than the virial mass, or a virial parameter ($\alpha_{vir}=M_{vir}/M_{clump}$) less than unity, indicates that the clump is gravitational unstable. We calculate the virial parameter of each clump and list the results in \autoref{tab:clump}. Except for the Clump 43 and 45, most of the clumps have $\alpha_{vir}$ less than unity, suggesting that thermal pressure and turbulence are insufficient to support the clump against gravity. Hence, these clumps require additional support from magnetic fields to stop gravitational collapse.

If support from magnetic fields is taken in to account, the virial mass of a clump becomes
\begin{equation}
M_{vir}^B=\frac{5R_{eff}}{G}(\sigma^2_{NT}+c_s^2+\frac{V_A^2}{6})
\end{equation}
\citep{be92,pi11,liu18}, where the additional term $\frac{V_A^2}{6}$ stands for the support from magnetic field pressure. We have estimated an Alfv\'{e}nic Mach number of $0.6\sin \theta$ for Clump 47 in \autoref{sec:str}, which corresponds to an Alfv\'{e}nic velocity of $0.64/\sin \theta$. With support from magnetic fields, the $\alpha_{vir}$ of Clump 47 becomes 0.2--1.0 for a $\theta$ of 15\degr--90\degr\, and greater than unity if $\theta<15\degr$. Hence, if the direction of magnetic field is not very close to the line of sight, Clump 47 is likely gravitationally unstable, which is consistent with the existence of YSOs in the central hub \citep{ha08}. In addition, the presence of YSOs in the Clump 47 indicates a density structure more complicate than our simple assumption, which could further reduce the virial mass \citep{sa17} and thus this clump could be even more unstable than our estimation.

\section{Discussion}\label{sec:discussion}
\subsection{The Origin of the Core-Scale HFS}\label{sec:HFS}
In \autoref{sec:results}, we show that the magnetic field orientation around the HFS have two major components, tending to distributed in the northern and southern part of the system. The two components can be explained by either a curved magnetic field or a foreground/background component overlaid on uniform magnetic field. Nevertheless, since the C$^{18}$O (J = 3-2) spectral data taken by \citet{gr08} shows that all components in the HFS are within a narrow velocities range ($\sim$ 3--5 km~s$^{-1}$), the first possibility is favored, unless the foreground/background component coincidentally has a velocity very similar to the HFS.

The curved magnetic field could be originated by an uniform large-scale magnetic field dragged by the contraction of the large-scale main filament. The dragging along the major axis of the large-scale filament would cause the single peak in large-scale magnetic field broadened and spilt into two peaks, and thus the center of the splitting shown in the histogram ($\sim 15\degr$) is consistent with the orientation perpendicular to the large-scale filament. In addition, the spatial distribution tendency of the two components can also be explained, since the contraction along the major axis would lead to an axisymmetric pattern. The feature that parsec-scale magnetic fields are perpendicular to filaments but modified by core collapsing in smaller scale has been found in other filamentary clouds, such as Serpens South \citep{su11}, Orion A \citep{pa17} and W51 \citep{ko18}.

Supercritical main filaments are expected to fragment along their major axis and trigger star formation \citep[e.g.,][]{an10,po11,mi12,an14,cl16}, which could be a possible origin of the observed HFS. The main filament connected to our observed HFS has a supercritical mass per unit length (152 M$_{\sun}$ ~pc$^{-1}$, \citealt{ar11}), and the submillimeter clumps identified in \citet{jo17} also indicate that some filament fragmentation has already taken place. 

Some theoretical work suggests that fragmentation of filaments with aspect ratios greater than 5 tends to first begin at their ends, where the edge-driven collapsing mode is more efficient than the homologous collapse mode over the whole filament \citep{po11,po12}. In contrast, the centralized collapse mode is more important in shorter filaments with high initial density perturbations or no magnetic support \citep{se15}. The edge-driven collapsing mode is consistent with the HFSs found at the end of the main filament in the IC5146 cloud. In addition, \citet{gr08} found the gas velocity within the main filament increases from the center to both ends, based on $^{13}$CO and C$^{18}$O line observations. The velocity gradient suggests that the gas within the main filament is likely fragmented towards the two massive HFSs at the ends. 

The center-to-ends filament fragmentation picture might seem inconsistent with the observed magnetic field morphology, which shows a pattern of end-to-center contraction. The curved magnetic field morphology, however, might be shaped at an early evolutionary stage, when the filament was still contracting and accumulating mass until its density was sufficiently high to trigger fragmentation. In addition, the global end-to-center contraction and the local center-to-end fragmentation could be occurring simultaneously but at different scales, as suggested by hierarchical gravitational fragmentation models \citep{go14,go18}.

\begin{figure}
\includegraphics[width=\columnwidth]{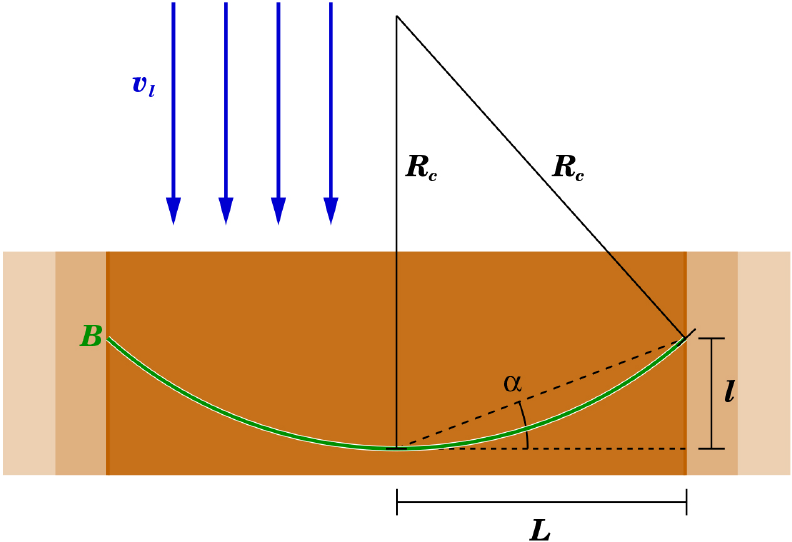}
\caption{Schematic for the magnetic field within an accretion flow, modeled in \citet{go18}. The magnetic field is bent by the ram pressure of the flow, and eventually reaches a stationary stage that the ram pressure balances the magnetic field tensor. The relative strength of the two forces determines the curvature radius and angle, $R_c$ and $\alpha$, by \autoref{eq:ushape}. This figure is adapted from the Figure 5 in \citet{go18} with permission.}\label{fig:go18}
\end{figure}

\begin{figure*}
\includegraphics[width=\textwidth]{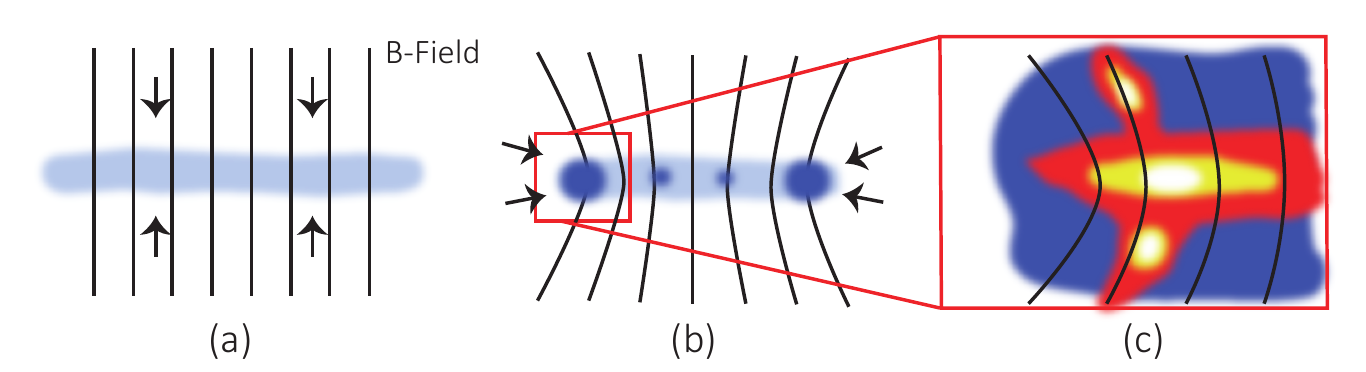}
\caption{A cartoon illustrates the possible formation scenario of the core-scale HFSs. (a) The parsec-scale filaments first form via the contraction and fragmentation along magnetic fields. (b) As the parsec-scale filaments become magnetically and thermally supercritical, the filaments fragment along their major axis, and the most massive components form at the end of filaments. At the same time, the magnetic fields are dragged by the filament contraction. (c) The massive fragment at the end of parsec-scale filament further fragments along the curved magnetic fields, and forms a core-scale HFS with orientation parallel or perpendicular to the local magnetic field instead of primordial field.}\label{fig:cartoon}
\end{figure*}

To explore the magnetic field morphology within collapsing clouds, \citet{go18} simulated molecular clouds undergoing global, multi-scale gravitational collapse. In this simulation, the magnetic fields would be dragged by the gravitational contraction, but eventually reach a stationary state in which the ram pressure of the flow balances the magnetic tension. Hence, the model predicts a random magnetic field morphology on parsec scales and a ``U-shape'' magnetic field within the filaments following the equation
\begin{equation}\label{eq:ushape}
\left(\frac{v_l}{v_A}\right)^2=2\sin(2\alpha),
\end{equation}
where $v_l$ is the gas velocity along the filaments, $v_A$ is the Alfv\'{e}nic velocity, and the $\alpha$ is the angle between the magnetic field line and the direction perpendicular to the filament, illustrated in \autoref{fig:go18}. Although the predicted large-scale random magnetic field morphology is inconsistent with the uniform magnetic fields shown by WLE17 data, a ``U-shape'' magnetic field within the filaments has been observed in our POL-2 data, suggesting that this model might be important when the filaments become dense enough. 

The observed $\alpha$ is $\sim$30\degr, estimated by the two components shown in the $PA$ histogram (\autoref{fig:hist_2comp}), so $v_l/v_A$ is expected to be 1.3 by \autoref{eq:ushape}. We assume that the $v_l$ is approximately the velocity difference along the filament around the central hub. The line of sight C$^{18}$O centroid velocity of the central hub (clump 47) is $\sim 4.1$ km~s$^{-1}$, and the western clump 42 has a centroid velocity of $\sim 3.8$ km~s$^{-1}$. Hence, the velocity difference along the filament between clumps 42 and 47 is $0.3/\cos \phi$ km~s$^{-1}$, where $\phi$ is the inclination angle of filament with respect to the line of sight. With the Alfv\'{e}nic velocity of $0.64/\sin \theta$ estimated in \autoref{sec:str}, the observed $v_l/v_A$ is $\sim 0.5 \sin \theta/\cos \phi$. Due to the unknown inclination angle, we can only speculate that the model expectation would be correct if the filament is nearly perpendicular to the line of sight ($\phi$ > 67\degr).

Based on our observed magnetic field morphologies, we propose a three-stage scenario to explain the origin of the observed HFS, illustrated in \autoref{fig:cartoon}. In the first stage, the large-scale magnetically subcritical filaments are first formed with dynamically important magnetic fields as described in strong magnetic field filament formation models \citep[e.g.,][]{na08,va14}, and these filaments appear perpendicular to a uniform large-scale magnetic field, as revealed by WLE17 data. In the second stage, the large-scale filaments accumulate mass via accretion along magnetic field lines or filament mergers \citep[e.g.,][]{li10, an14}, and eventually become magnetically and thermally supercritical. The contraction of supercritical filaments would bend the uniform primordial magnetic fields, similar to the case in Orion A \citep{pa17}. In the third stage, the dense clumps within filaments, often at the ends of filaments, would tend to fragment along magnetic fields and form second generation filaments with hub-filament morphologies, because density perturbations parallel to the magnetic fields grow faster than those perpendicular to the fields\citep[e.g.,][]{na98,va14}. The collapse of the cores within the second generation filaments is also regulated by the bent magnetic fields, and so the cores are oriented either parallel or perpendicular to local magnetic fields, as shown in \autoref{fig:papa}, and are less correlated to the primordial magnetic field.

\subsection{The Alignment between Local Magnetic Fields and Clumps}\label{sec:alignclump}
Stars form predominantly from high column density filaments \citep{an10}. Although most filaments are either oriented parallel or perpendicular to the large-scale magnetic fields \citep{li14,pl16}, only a few young stars have been observed with hourglass magnetic field morphologies which favors a star formation scenario where the core collapse is regulated by strong magnetic fields \citep[e.g.,][]{gi06,ra09,ta09}. As a counterexample, ALMA polarization observations toward the embedded source Ser-emb 8 show chaotic magnetic fields \citep{hu17}, indicating that this star was formed under weak magnetic field conditions. This difference poses the question of how physical scales and environments generally determine the role of magnetic fields in star formation. 

To address the role of magnetic fields in star formation, the SMA polarization survey toward massive cores \citep{zh14} revealed that magnetic fields on the core scale (0.1--0.01 pc) are mostly either parallel or perpendicular to the magnetic fields on the parsec scales. \citet{li15} further analyzed the magnetic field morphologies in NGC 6334 on the 100 pc to 0.01 pc scales, and found that local magnetic fields on all these scales are either parallel or perpendicular to the local cloud elongation. Both these results suggest that magnetic fields are dynamically important during the collapse and fragmentation of clouds, possibly guiding the contraction of filaments and cores. \citet{ko14} further used a large sample set (50 sources) to examine the bimodal distribution of the relative orientation between the magnetic fields and the density structures, and found that the distribution is more scattered than those in previous surveys, although a bimodal distribution cannot be ruled out.

In \autoref{sec:clump}, we find a tendency that the clumps within the observed HFS have orientations parallel or perpendicular to the local magnetic fields (at 0.1--0.01 pc scale). The local magnetic fields in many of these clumps, however, have orientations 30--60\degr\ different from the parsec scale magnetic field, which is inconsistent with the findings of \citet{zh14} and \citet{li15}. The inconsistent cases are mainly clumps within the extending filaments, which follow the orientation of the curved magnetic fields (see \autoref{sec:HFS}). These clumps are much fainter than those in the central hub, which possibly explains why they were missed in previous surveys. Nevertheless, since we still found the orientations of these clumps to be well coupled with the host filaments and the local magnetic fields, our results support the idea that magnetic fields are important in regulating the core and filament collapse on spatial scales of 0.01-0.1 pc.

\section{Conclusions}\label{sec:summary}
This paper presents the first-look results of SCUBA-2/POL-2 observations at 850 $\mu$m towards the IC5146 filamentary clouds as part of the BISTRO project. Our observations reveal the magnetic field morphology within a core-scale Hub-Filament Structures (HFS) located at the end of a parsec-scale filament. From the analysis of these data, we find:
\begin{enumerate}
\item The observed polarization fraction is found to vary with Stokes $I$ following a power-law with an index of $\approx0.56$, which suggests that dust grains in this $A_V$ $\sim$20--300 mag range can be still aligned with magnetic fields.

\item The observed polarization map shows that the magnetic field of the HFS on core-scales ($\sim 0.05$--$0.5$ pc) is more organized than random. The core-scale magnetic field is likely inherited from a larger scale magnetic field that has been dragged by contraction along the parsec-scale filament.

\item The submillimeter clumps within the observed core-scale HFS tend to be aligned with local magnetic fields, i.e., they are oriented within 20\degr\ of being either parallel or perpendicular to the local magnetic field direction. This trend may suggest that the core-scale HFS formed after the parsec-scale filament became magnetically supercritical, and the magnetic fields have been dynamically important during the formation and the following evolution of the core-scale HFS.

\item We propose a scenario to explain the formation of the core-scale HFS: the parsec-scale filaments first form under a strong and uniform magnetic field, and started to fragment and locally bend the magnetic field as they becomes magnetically supercritical. The massive clump, formed at the end of the parsec-scale filament, further fragments under the strong magnetic fields and becomes the core-scale HFS.

\item Using the Davis-Chandrasekhar-Fermi method, the magnetic strength within the central hub is estimated to be $0.5\pm0.2$ mG, and the mass-to-flux ratio is $1.3\pm0.4$ for D = 813 pc. The Alfv\'{e}nic Mach number estimated using the magnetic field angular dispersion is $<$0.6. These estimates suggests that gravity and magnetic fields is comparably important in the current core-scale HFS, and that turbulence is less important.

\end{enumerate}

\acknowledgments
The James Clerk Maxwell Telescope is operated by the East Asian Observatory on behalf of The National Astronomical Observatory of Japan; Academia Sinica Institute of Astronomy and Astrophysics; the Korea Astronomy and Space Science Institute; the Operation, Maintenance and Upgrading Fund for Astronomical Telescopes and Facility Instruments, budgeted from the Ministry of Finance (MOF) of China and administrated by the Chinese Academy of Sciences (CAS), as well as the National Key R\&D Program of China (No. 2017YFA0402700). Additional funding support is provided by the Science and Technology Facilities Council of the United Kingdom and participating universities in the United Kingdom and Canada. Additional funds for the construction of SCUBA-2 and POL-2 were provided by the Canada Foundation for Innovation. The Starlink software \citep{cu14} is currently supported by the East Asian Observatory. JCMT Project code M16AL004. The authors wish to recognize and acknowledge the very significant cultural role and reverence that the summit of Maunakea has always had within the indigenous Hawaiian community. We are most fortunate to have the opportunity to conduct observations from this mountain. J.W.W., S.P.L., K. P., and C.E. are thankful to the support from the Ministry of Science and Technology (MOST) of Taiwan through the grants MOST 105-2119-M-007-021-MY3, 105-2119-M-007-024, and 106-2119-M-007-021-MY3. J.W.W. is a University Consortium of ALMA--Taiwan (UCAT) graduate fellow supported by the Ministry of Science and Technology (MOST) of Taiwan through the grants MOST 105-2119-M-007-022-MY3. M.K. was supported by Basic Science Research Program through the National Research Foundation of Korea (NRF) funded by the Ministry of Science, ICT \& Future Planning (No. NRF-2015R1C1A1A01052160). C.W.L. is supported by the Basic Science Research Program through the National Research Foundation of Korea (NRF), funded by the Ministry of Education, Science and Technology (NRF-2016R1A2B4012593). Woojin Kwon was supported by Basic Science Research Program through the National Research Foundation of Korea (NRF-2016R1C1B2013642). D. J. is supported by the National Research Council of Canada and by an NSERC Discovery Grant.

\appendix
\section{Bias on determination of $P$ vs. $I$ Relation}\label{sec:appendix}
In this section, we use the simulated data to investigate the two possible sources of bias in the estimation of the $P$--$I$ relation. The first bias source is the selection criteria $P/\delta P>2$. Since the uncertainty in the polarization fraction has the following dependence on total intensity of 
\begin{equation}
\delta P\approx \frac{\delta Q}{I} \propto I^{-1},
\end{equation}
assuming $\delta Q \approx \delta U$, the selection criteria would truncate the observed $P$--$I$ distribution by a boundary of $P=2\delta P\propto I^{-1}$. Hence, the selection criteria could cause an artificial trend of $P\propto I^{-1}$ to arise, leading to the erroneous conclusion that the dust cannot be aligned with magnetic fields. This bias source could be avoided simply by including the low SNR data.

The second bias source is the non-Gaussian PDF of the observed polarization fraction. As shown in \autoref{sec:pext}, the PDF of the observed polarization fraction follows the Rice distribution (\autoref{eq:rice}). The Rice distribution can be approximated as a Gaussian when $P/\delta P \gtrapprox 4$, but it becomes more asymmetric as $P/\delta P$ decreases \citep{va06}. The misuse of a Gaussian PDF on Rice distributed data would cause the polarization fraction to be overestimated. Furthermore, the bias is anticorrelated with the SNR of $P$, as well as $I$, and steepen the measured slope. The commonly used debiasing methods, e.g., the asymptotic estimator, however, could help remove the bias in the high $P/\delta P$ domain, but the PDF of the debiased polarization fraction would be still non-Gaussian \citep{mo15b}. To avoid this bias source, using an appropriate PDF, instead of a Gaussian assumption, to analyze the observed polarization fraction is recommended.

\begin{figure*}
\includegraphics[width=\textwidth]{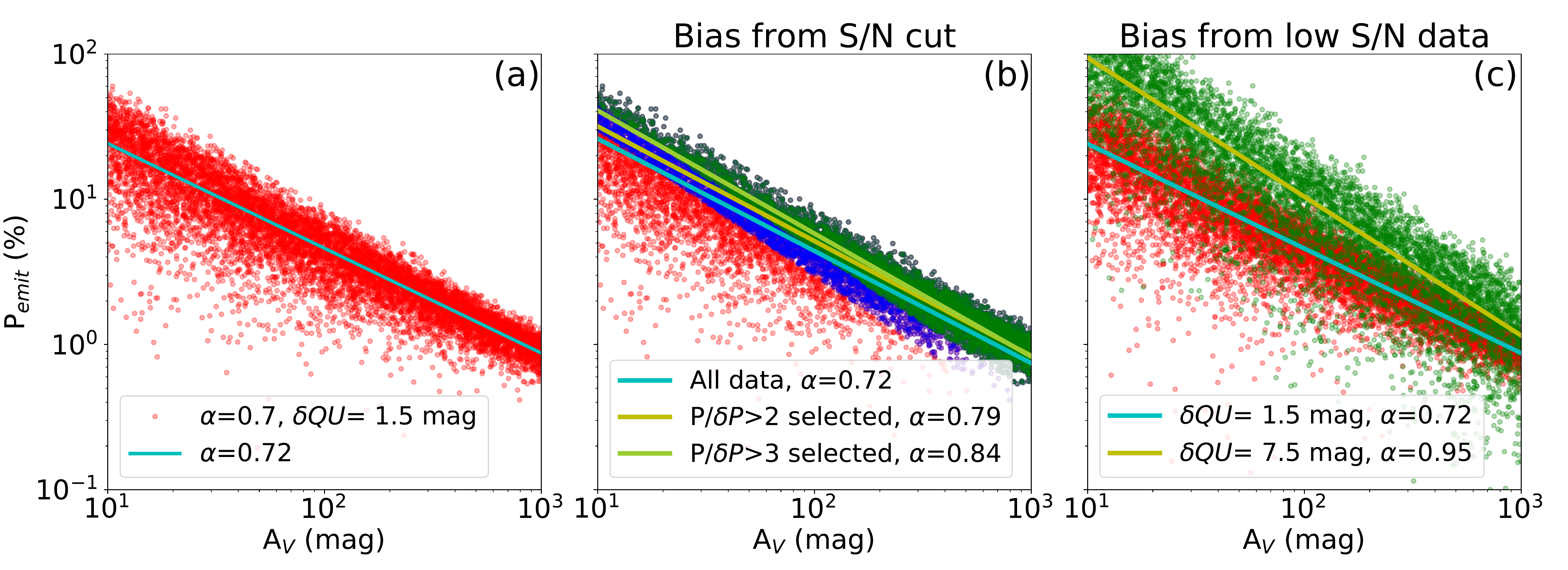}
\caption{Monte Carlo simulations to show the possible bias for fitting the $P$--$I$ relation using P$\propto I^{-\alpha}$. (a) The simulated data assuming P$\propto I^{-0.7}$ with moderate noise of 1.5 mag in Stokes $Q$ and $U$. The least-squared fitting could recover the $\alpha$ of $\sim 0.7$. (b) Sample selection criteria of $P/\delta P>2$ (blue) and $P/\delta P>3$ (green) were applied to the simulated data. The derived $\alpha$ of the selected data using least-squared fitting become much higher than input model. (c) The green points represent the simulated data with 5 times higher noise. The derived $\alpha$ for the high noise set using least-squared fitting is almost 1.}\label{fig:simP}
\end{figure*}

We performed Monte Carlo simulations to generate a set of polarization data with a given $P$--$A_V$ relation, and tested how the measured relation is affected by the bias sources discussed above. To simulate the observed polarization data, we first generated a 10000-element set of $A_V$ values distributed uniformly in logarithm where the polarization fraction of each element was determined by the underlying power-law 
\begin{equation}
P\propto A_V^{-0.7}.
\end{equation}
For simplicity, we assumed all polarization vectors have $PA$s of 0\degr, and we calculated the Stokes $Q$ and $U$ values for each pair of ($A_V$, $P$). Here we directly used $A_V$ magnitude as the intensity unit. We also added Gaussian-distributed noise with $\sigma = 1.5$ $A_V$ mag to both Stokes $Q$ and $U$, and calculated the debiased polarization fraction from the noise-included Stokes $Q$ and $U$. The simulated $P$ vs. $A_V$ distribution is plotted in \autoref{fig:simP} (a). A least-squares power-law fit to the simulated data returns an index of $0.725\pm0.002$, similar to our input value.

To investigate how the first bias source (selection criteria) affects the $P$--$A_V$ relation, we applied selection criteria $P$/$\delta P > 2$ and $P$/$\delta P > 3$ to the simulated data, as shown in \autoref{fig:simP} (b). It is clearly shown that the $P$--$A_V$ distribution is trunctaed by the applied selection criteria, and the least-squared fits here return power-law indices of $0.796\pm0.002$ and $0.850\pm0.003$ for the data selected by $P$/$\delta P > 2$ and $P$/$\delta P > 3$, respectively, suggesting that such selection criteria would significantly steepen the measured slope.

To test how the second bias source (non-Gaussianity when $P/\delta P$ is low) affects the $P$--$A_V$ relation, we generated another simulated data set with five times higher noise in Stokes $Q$ and $U$ than our original data set. \autoref{fig:simP} (c) shows the $P$--$A_V$ distribution from both the simulated data with original and higher noise. The simulated data with higher noise show a steeper trend, which mainly results from the positive bias in the low $A_V$ or low $P/\delta P$ regimes. A power-law index of $0.950\pm0.006$ for the high noise set was obtained by the least-squared fitting, significantly higher than the $0.725\pm0.002$ derived from the original set. These examples show that the two bias sources both could steepen the measured $P$--$I$ relation, we would further explore how significant the effects are in a much wider parameter space in our following paper \citet{pa19}.

\begin{figure*}
\includegraphics[width=\textwidth]{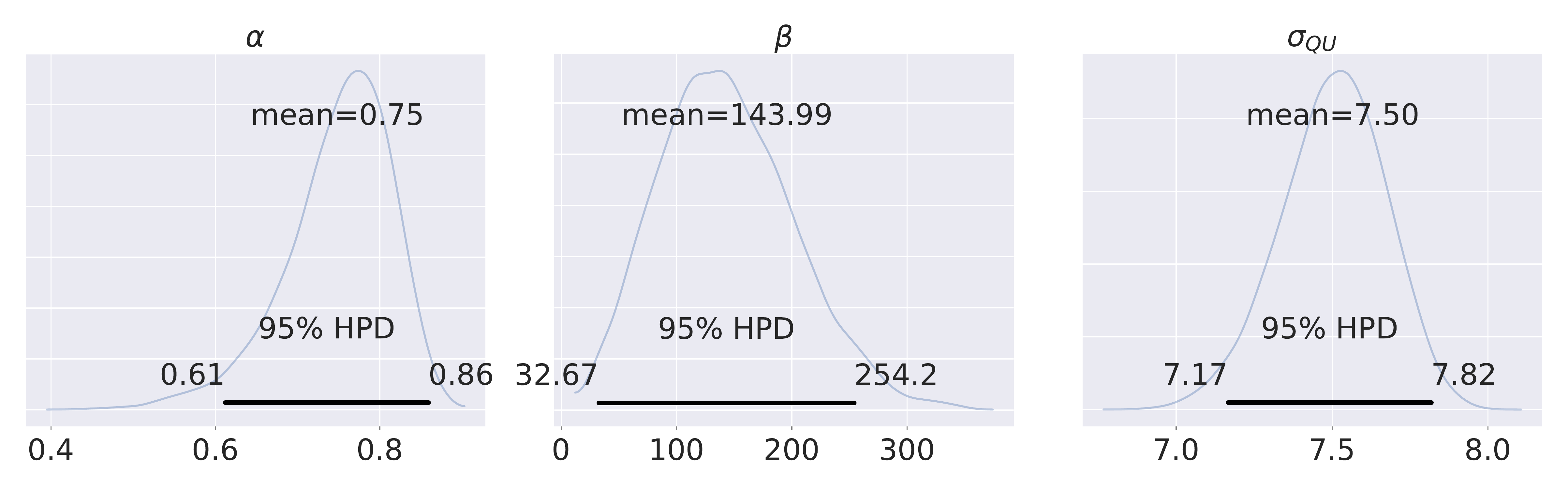}
\caption{The PDF of model parameters obtained from the Bayesian fitting to the simulated data with high noise (\autoref{fig:simP} green points). The mean values and 95\% highest posterior density (HPD) regions are labeled in each panel. All the results are consistent with our input model.}\label{fig:bayesian_test}
\end{figure*}

To test whether or not our Bayesian model, as described in \autoref{sec:pext}, can avoid the bias due to the non-Gaussianity. We fit the same higher noise simulated data set with our Bayesian model. The derived PDFs of each model parameter are shown in \autoref{fig:bayesian_test}. All the derived model parameters are consistent with our input values, and the mean posterior suggests an $\alpha$ of 0.75, which is much more accurate than the $\alpha$ of 0.95 derived from least-squared fitting. In addition, the possibility of an asymmetric PDF is considered in the Bayesian model, and hence it provides a more realistic uncertainty estimation. 

\facility{JCMT}
\software{Aplpy \citep{ro12}, Astropy \citep{as13}, NumPy \citep{va11}, PyMC3 \citep{sal16}, PySpecKit \citep{gi11}, SciPy \citep{jo01}, Starlink \citep{cu14}}

\end{document}